\documentclass[a4paper,11pt]{article}
\usepackage{aaskaiid}
\usepackage{orcidlink}
\usepackage{hyperref}
\usepackage{cleveref}
\usepackage{multirow}

\usepackage{xspace}

\newcommand*{\mr}[1]{{\mathrm{#1}}}

\newcommand*{\updated}[1]{\textcolor{black}{#1}\xspace}

\renewcommand{\ne}{n_{\rm e}}
\newcommand{\Blos}{{B_{\parallel}}}
\newcommand{\rad}{{\rm rad}}
\newcommand{\DM}{{\rm DM}}
\newcommand{\RM}{{\rm RM}}
\newcommand{\m}{{\rm m}}
\newcommand{\cm}{{\rm cm}}
\newcommand{\dis}{d_{\rm FRB}}
\newcommand{\pc}{{\rm pc}}
\newcommand{\kpc}{{\rm kpc}}
\newcommand{\muG}{\mu{\rm G}}
\newcommand{\dd}{{\rm d}}
\renewcommand\vec[1]{{\mathbf{#1}}}
\newcommand{\Ningal}{{N_{\rm ingal}}}
\newcommand{\ingal}{{{\rm ingal}}}
\newcommand\Sec[1]{Sec.~\ref{#1}}
\newcommand\Tab[1]{Table~\ref{#1}}

\newcommand{\hubble}{$H_0$}

\title{Fast Radio Bursts as Cosmological Probes}
\ShortTitle{FRBs: Cosmological Probes}

\ShortName{Caleb et al.} % shortened name list for header 
\author[1]{Manisha Caleb\orcidlink{0000-0002-4079-4648}}
\author[2,3]{Jéferson A. S. Fortunato\orcidlink{0000-0001-7983-1891}}
\author[4,5]{Steffen Hagstotz\orcidlink{0000-0002-7044-3793}}  
\author[6]{Clancy W. James\orcidlink{0000-0002-6437-6176}}
\author[7,8]{Joscha N. Jahns-Schindler\orcidlink{0000-0003-4193-6158}}
\author[9]{Dylan Jow\orcidlink{0000-0003-3236-8769}}
\author[10]{Evan Keane\orcidlink{0000-0002-4553-655X}}
\author[11]{Koustav Konar\orcidlink{0000-0002-8236-1605}}
\author[11]{Yin-Zhe Ma\orcidlink{0000-0001-8108-0986}}
\author[12]{Daniele Michilli\orcidlink{0000-0002-2551-7554}}
\author[13]{Robert Reischke$^*$\orcidlink{0000-0001-5404-8753}}
\author[14]{Amit Seta\orcidlink{0000-0001-9708-0286}}
\author[15]{Priyanka Singh$^*$}
\author[16]{Laura G. Spitler$^*$\orcidlink{0000-0002-3775-8291}}
\author[17]{Yidan Wang\orcidlink{0000-0002-7372-4160}} 
\author[2,3]{Amanda Weltman\orcidlink{0000-0002-5974-4114}}
\author[]{The SKA Transients SWG}

\affiliation[1]{Sydney Institute for Astronomy, School of Physics, The University of Sydney, Sydney, NSW 2006, Australia}
\affiliation[2]{High Energy Physics, Cosmology \& Astrophysics Theory (HEPCAT) Group, Department of Mathematics and Applied Mathematics, University of Cape Town, Cape Town, 7700, South Africa}
\affiliation[3]{African Institute for Mathematical Sciences, 6 Melrose Road, Muizenberg, Cape Town, 7945, South Africa}
\affiliation[4]{Universit\"ats-Sternwarte, Fakult\"at f\"ur Physik, Ludwig-Maximilians Universit\"at M\"unchen, Scheinerstraße 1, D-81679 M\"unchen, Germany and}
\affiliation[5]{
Excellence Cluster ORIGINS, Boltzmannstraße 2, D-85748 Garching, Germany}
\affiliation[6]{International Centre for Radio Astronomy Research, Curtin University, Bentley, 6102, WA, Australia}
\affiliation[7]{Centre for Astrophysics and Supercomputing, Swinburne University of Technology, Hawthorn, VIC 3122, Australia}
\affiliation[8]{ARC Centre of Excellence for Gravitational Wave Discovery (OzGrav), Hawthorn, VIC 3122, Australia}
\affiliation[9]{Kavli Institute for Particle Astrophysics and Cosmology, Stanford University, 452 Lomita Mall, Stanford, CA 94305, USA}
\affiliation[10]{School of Physics, Trinity College Dublin, College Green, Dublin 2, D02 PN40, Ireland}
\affiliation[11]{Department of Physics, Stellenbosch University, Matieland 7602, South Africa}
\affiliation[12]{Laboratoire d’Astrophysique de Marseille, Aix-Marseille University, CNRS, CNES, Marseille, France}
\affiliation[13]{Argelander-Institut f\"ur Astronomie, Universit\"at Bonn, Auf dem H\"ugel 71, D-53121 Bonn, Germany}
\affiliation[14]{Research School of Astronomy and Astrophysics, Australian National University, Canberra, ACT 2611, Australia}
\affiliation[15]{Department of Astronomy, Astrophysics and Space Engineering, Indian Institute of Technology, Indore 453552, India}
\affiliation[16]{Max-Planck-Institut f\"ur Radioastronomie, Auf dem H\"ugel 69, 53121 Bonn, Germany}
\affiliation[17]{National Astronomical Observatories, Chinese Academy of Sciences, Beijing 100101, China University of Chinese Academy of Sciences, Beijing 100049, China}
\affiliation[*]{Coordinating author}
\emailAdd{reischke@posteo.net}
\emailAdd{lspitler@mpifr-bonn.mpg.de}

\abstract{Fast radio bursts (FRBs) are brief, coherent radio pulses of extragalactic origin. They typically last from microseconds to milliseconds and have energies large enough to be visible over cosmological distances. Since FRBs interact with free electrons along their paths, the original burst is dispersed (Dispersion Measure, DM) and broadened (scattering). Furthermore, the burst's polarization is altered by Faraday rotation. Consequently, FRBs are excellent probes of the cosmological distribution of baryons, the expansion of the Universe, magnetic fields, and minuscule effects of fundamental physics that accumulate over vast distances. 

This chapter is the second of a trilogy of FRB chapters and discusses FRBs as a standalone probe.
We first introduce the foundation of FRB observables related to those questions. Next, we lay the groundwork for forecasting SKA’s potential by describing the method to simulate the expected FRB population observable with the SKA. These synthetic FRB catalogues are then used to investigate the SKA's potential to probe the Universe’s expansion rate and fundamental physics, such as the equivalence principle and the existence of massive photons. Furthermore, we investigate the possibility of tracing cosmic magnetic fields and investigating different dark matter candidates.}

\begin{document}
\maketitle

\section{Introduction}

Fast Radio Bursts (FRBs) are the first extragalactic radio sources with temporal variation on $\sim$ms timescales \citep[see][for a review]{Petroff2019}, and due to their large luminosities, can be detected out to distances of many Gpc \citep{2025arXiv250801648C}. The physical properties of the ionized media along the FRB's path are encoded in the radio signal through several propagation effects, most prominently dispersion. They are also abundant, with $>10^5$ occurring over the sky per day \citep[e.g.][]{crafts21}. Therefore, FRBs are a unique probe of the Universe.

The search for new FRBs with SKA is well suited for commensal observations with other surveys. Most cosmological applications benefit from a large number of sightlines, but their exact locations on the sky is generally not important. One exception is FRBs that travel through the Galactic plane may be of lower value for cosmology, because the Galaxy makes optical follow-up of the host galaxy difficult and imposes additional dispersion that can have high uncertainty, but provides a new probe of the Milky Way interstellar medium. In summary, an FRB search should be running whenever and wherever SKA is observing in order to detect as many FRBs over as much of the sky as possible.

Compared to other big telescopes in the FRB game, SKA excels in sensitivity and frequency coverage.
Current telescopes that lead the blind search for FRBs are ASKAP \citep{Hotan2021,Shannon2024}, CHIME \citep{CHIME2021, CHIME2022}, DSA-110\footnote{The telescope names are acronyms standing, respectively, for Australian SKA Pathfinder, Canadian Hydrogen Intensity Mapping Experiment, and Deep Synoptic Array} \citep{Law2023}, and MeerKAT \citep[e.g.][]{Rajwade+22,caleb_2023_subarcsec,2025arXiv250801648C}. 
SKA-Mid will have higher sensitivity than these telescopes, but a lower instantaneous field of view in some cases.
As a result, the expected detection rate is comparable to CHIME's current rate, although the FRB population observed by SKA will have a higher average redshift.
FRBs at high redshifts are of particular interest, as we will discuss further below, and SKA, with its high sensitivity, will discover many of them.

In this chapter we provide the necessary background information for the use of FRBs as cosmological probes and describe a simulated sample of FRBs that could be detected by SKA-mid bands 1 and 2 and SKA-low for the AA$^*$ and AA4 configurations, based on our current knowledge of the population. These simulations are used as input for quantitative forecasts of several cosmological applications of FRBs. In this chapter we focus on the science of extragalactic magnetic fields (\ref{sec:mag}), the expansion of the Universe (\ref{sec:expansion}), dark matter searches (\ref{sec:dark}), and fundamental physics (\ref{sec:fundi}). A companion chapter focuses on using FRBs to study the distribution of baryonic matter in the Universe \citep{Caleb01.2026.SKA}. In addition, a separate chapter describes the potential of SKA for understanding the central engine(s) of FRBs \citep{Curtin01.2026.SKA}.

\begin{figure}
    \centering
 \includegraphics[width=0.45\linewidth]{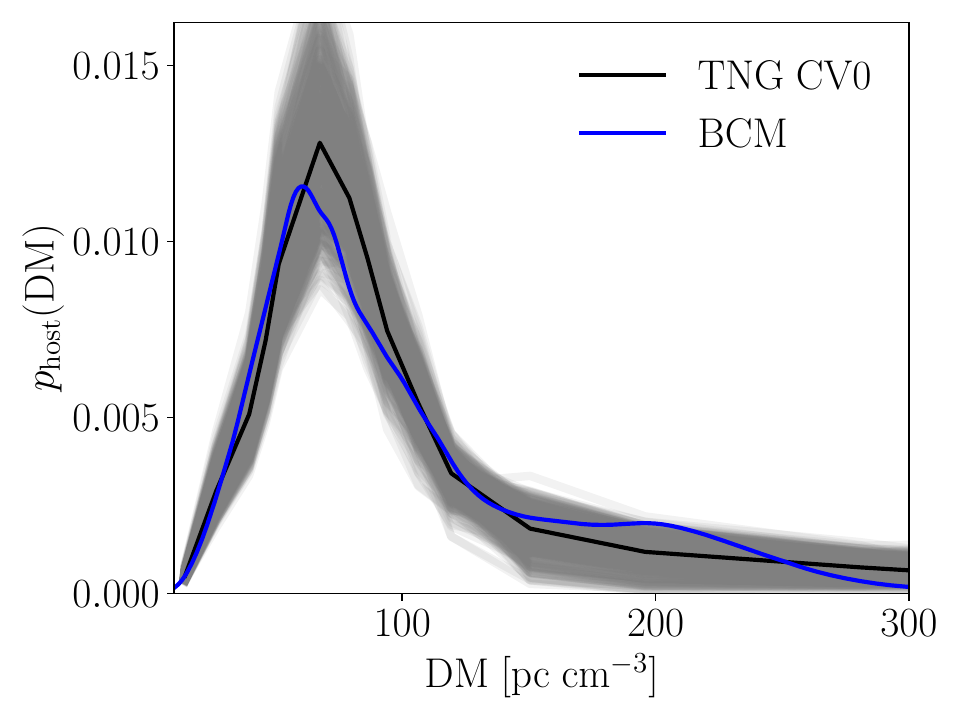}
\includegraphics[width=0.45\linewidth]{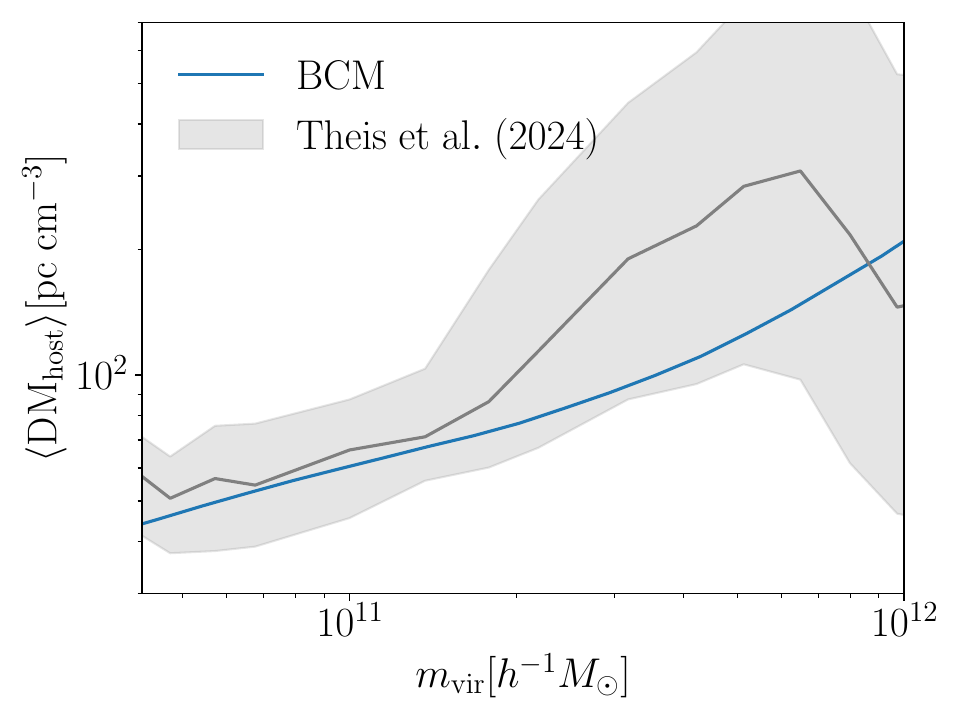}
    \caption{\textit{Left}: Typical distributions of the host contribution as measured in the TNG simulation by \citet{2024arXiv240308611T} with the model prediction from \citet{2025OJAp....8E.127R}. \textit{Right}: dependence of the host contribution on halo mass. }
    \label{fig:host}
\end{figure}

\section{FRB Observables}
In this section, we describe the fundamental observables that are used to analyse the medium traversed by FRBs. We will go into more detail about the different contributions in the relevant sections in this chapter, as well as in the FRBs as baryon tracers chapter \citep{Caleb01.2026.SKA}.

\subsection{The dispersion measure}
The dominating effect in FRB observations is the dispersion measure (DM). The arrival time of the photons at frequency $\nu$ is an integral over the inverse of the burst's group velocity. Using the dispersion relation of electromagnetic waves propagating through a cold plasma and expanding the inverse group velocity in terms of plasma frequency, one finds that the time delay is
    $\Delta t \propto \nu^{-2}$ and one defines the proportionality constant to be the DM
    \begin{equation}
    \label{eq:dm_definition}
    \mathrm{DM } = \int_0^{d_\mathrm{FRB}} n_\mathrm{e}(s)\,\mathrm{d}s\;,
\end{equation}
where $d_\mathrm{FRB}$ is the distance to the FRB, $n_\mathrm{e}(s)$ is the free electron density, and the integral is carried out along the line-of-sight. Observationally, the DM of an FRB at redshift $z$ (inferred from $d_\mathrm{FRB}$ or vice versa) and (two-dimensional) sky position $\boldsymbol{x}$ consists of different contributions from the distinct environments encountered by the burst:
\begin{equation}
\label{eq:parts}
    \mathrm{DM}_\mathrm{obs}(\mathbf{x},z) = \mathrm{DM}_{\mathrm{host}}(z) + \mathrm{DM}_{\mathrm{cosmic}} (\mathbf{x},z)+ \mathrm{DM}_{\mathrm{MW}}(\mathbf{x}) + \mathrm{DM}_{\mathrm{MW,halo}}(\mathbf{x})\,.
\end{equation}
$\mathrm{DM}_{\mathrm{host}}$ is the host contribution originating from the burst's host galaxy as well as from its local environment. $\mathrm{DM}_{\mathrm{cosmic}}$ describes the contribution from the mean free electron density and the large-scale structure (LSS) in the Universe. Lastly,  $\mathrm{DM}_{\mathrm{MW}}(\mathbf{x})$ and $ \mathrm{DM}_{\mathrm{MW,halo}}(\mathbf{x})$ are the contributions from the Milky Way and its halo. Hence, analysing the observed DM requires modelling each component. However, as indicated in Eq.~(\ref{eq:parts}), observing FRBs along different sightlines and at different redshifts can statistically distinguish the contributions. The following provides a short summary of the different contributions. A more detailed discussion will be provided in the corresponding sections.

\subsubsection{The host contribution}
The host contribution appears to be drawn from a highly skewed distribution with a long tail to large DM as a result of FRBs going off in the trailing edge of their respective host galaxy. This is also supported by analytical arguments \citep{mcquinn_locating_2014,2025OJAp....8E.127R} and directly from hydrodynamic simulations \citep{2020AcA....70...87J,2024ApJ...967...32M,2024arXiv240308611T}. There is no agreement on whether there is a functional form to describe the probability distribution function of the host. However, a log-normal distribution can describe current data sufficiently well. In \Cref{fig:host}, we show the typical form of the host contribution as found in simulation and its dependence on mass.

\subsubsection{The Milky Way contribution}
\begin{figure}
    \centering    \includegraphics[width=7cm, height=4.6cm]{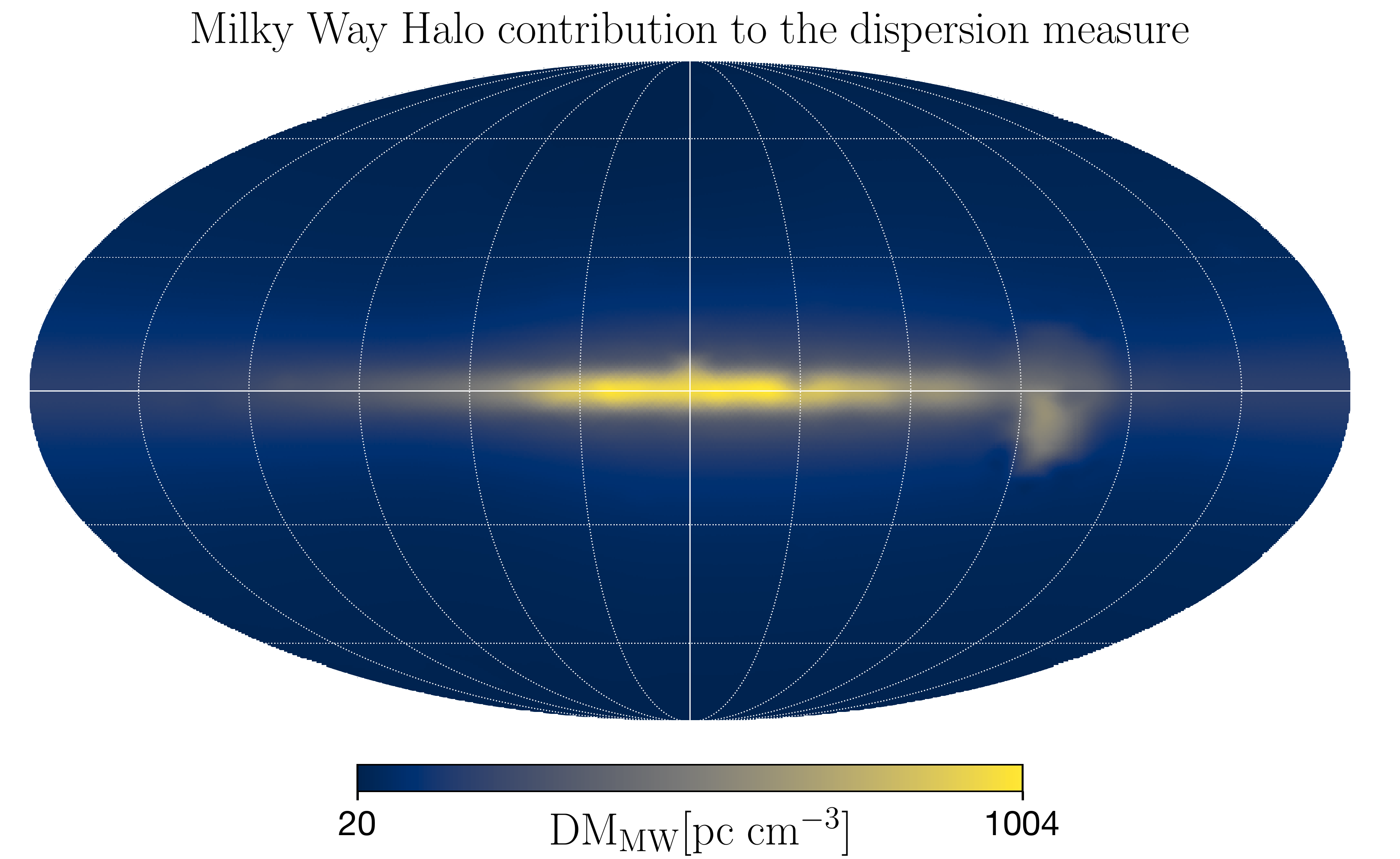}
\includegraphics[width=7cm, height=4.6cm]{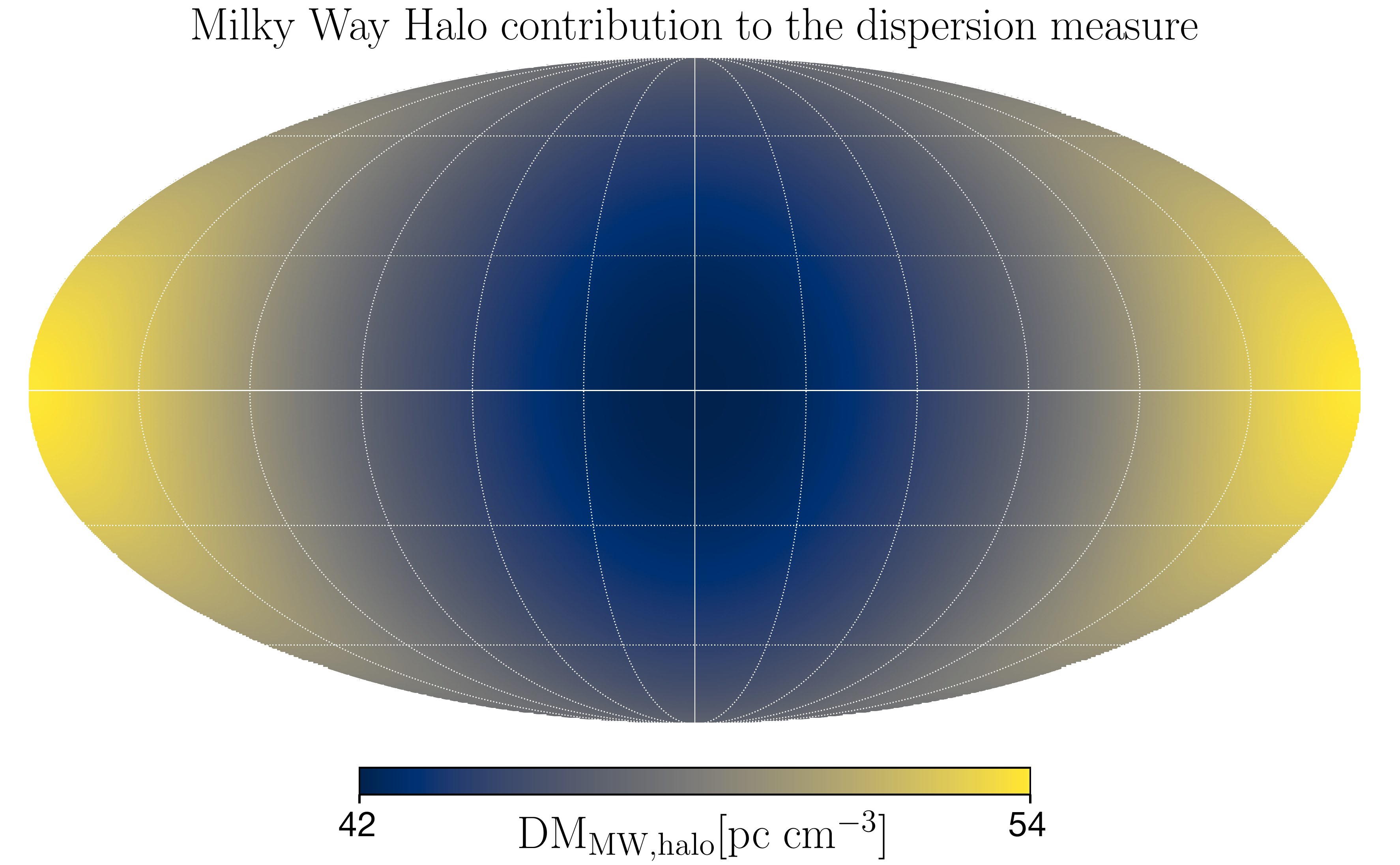}
    \caption{\textit{Left}: Contribution of the MW to the DM, calculated using the code from \citet{2024RNAAS...8...17O}. \textit{Right}: Possible contribution to the DM from the MW halo, shown in galactic coordinates. The profile is derived using the model presented in \citet{2025OJAp....8E.127R} for a halo $M= 1.5\times 10^{12}M_\odot$.}
    \label{fig:mw_halo_contribution_spherical}
\end{figure}

The distinction of the MW contribution is useful because $\mathrm{DM}_\mathrm{MW}$ can be modeled quite accurately at high galactic latitudes \citep{2017ApJ...835...29Y,2020ApJ...897..124O,2023MNRAS.523.6264R} using pulsar DMs. $\mathrm{DM}_\mathrm{MW,halo}$, however, is less well constrained and can vary significantly \citep[e.g.][]{2019MNRAS.485..648P,2020ApJ...888..105Y,2023ApJ...946...58C}, with values reaching up to $100\,\mathrm{pc}\,\mathrm{cm}^{-3}$, these are, though, still debated. To first order, one can of course model this contribution by a spherically symmetric halo \citep[see e.g.][]{2020MNRAS.496L.106K}. In \Cref{fig:mw_halo_contribution_spherical} we show the contribution of the MW and the MW halo. For the former we use the model by \citet{2024RNAAS...8...17O} and the MW halo contribution is calculated with the model described in \citet{2025OJAp....8E.127R}. 

One can see that at low galactic latitude the disk dominates and therefore produces large DM. For $l>30$, however, the halo contribution can become more important. Understanding both contributions will become very important to unlocking FRB science. Here, the SKA can deliver a unique view into the southern hemisphere.

\subsubsection{The cosmological contribution}
$\mathrm{DM}_\mathrm{cosmic}$ inherits the statistical properties of the LSS. Using Eq. (\ref{eq:dm_definition}), one can express it as follows:
\begin{equation}
\label{eq:DM_LSS_v2}
    \mathrm{DM}^{\mathstrut}_\mathrm{cosmic}({\boldsymbol{x}},z) = \frac{3 \Omega_\mathrm{b0} \chi_\mathrm{H}}{8 \uppi G m_\mathrm{p}}\chi_\mathrm{e}  \int_0^z   \frac{1+z'}{E(z')} f^{\mathstrut} (z^\prime) \big[1+\delta^{\mathstrut}_\mathrm{e}({\boldsymbol{x}},z')\big] \mathrm{d} z' ,
\end{equation}
with the dimensionless baryon density parameter $\Omega_\mathrm{b0}$ today, the Hubble radius, $\chi_\mathrm{H} = c/H_0$, the proton mass $m_\mathrm{p}$, the number of electrons per baryon $\chi_\mathrm{e}$ and the gravitational constant $G$. $\delta_\mathrm{e}$ is the electron density contrast, $E(z)$ is the expansion function, and $f(z)$ is the fraction of ionised baryons, as only these electrons contribute to the DM. The average of Eq. (\ref{eq:DM_LSS_v2}) yields the well-known Macquart relation \citep{macquart_census_2020}.

\subsection{Scattering}

Scattering arises due to the multi-path propagation of the burst through an ionized medium. Inhomogeneities in the electron density transverse to the line of sight lead to a frequency-dependent deflection in the rays:
\begin{equation}
    \alpha = \frac{\lambda^2 r_e}{2\pi} \int \nabla_\perp n_e(\mathbf{x}, s)\, \mathrm{d}s,
    \label{eq:scattering_angle}
\end{equation}
where $\alpha$ is the bending angle, $\lambda$ is the wavelength at the scatterer, and $\nabla_\perp$ refers to the gradient in the coordinate transverse to the line of sight, $\mathbf{x}$. When the bending angle is sufficiently large, multiple rays may connect the source to the observer, with offset times of arrival. While the term `plasma lensing' describes this phenomenon more generally, the term `scattering' is typically used to refer to situations where there are a large number of such rays and observable quantities are computed statistically. The primary observable is the scattering timescale, or the mean time delay of the deflected rays relative to the direct line of sight, $\tau_{\rm scatt.} = \frac{1+z}{2c}\frac{d_\mr{sl} d_\mr{lo}}{d_\mr{so}} \langle \alpha^2 \rangle$, where $d_\mr{sl}, d_\mr{lo},$ and $d_\mr{so}$ are the (angular diameter) distances between the source and scatterer, scatterer and observer, and source and observer, respectively. When the scattering time is large relative to the intrinsic burst width, scattering manifests as an exponential `scattering tail' in the burst with a frequency-dependent width. The width of this tail is measured to determine the scattering time. When the scattering time is small relative to the intrinsic burst width, the images interfere at the observer, leading to a frequency-dependent scintillation pattern. The `scintillation bandwidth' is measured and is related to the scattering time via $\Delta \nu_{\rm scint} = 2 \pi / \tau_{\rm scatt.}$ \citep{Pradeep2025}. 

The frequency dependence of these quantities can be used to confirm the scattering origin of the observed effect, with $\tau_{\rm scatt.} \propto \nu^{-\delta}$, where $\delta \sim [4, 4.4]$. The scaling $\delta = 4$ arises when the scattering is purely geometric (for which Eq.~\ref{eq:scattering_angle} holds) and $\delta = 4.4$ when the scattering is purely diffractive (i.e. wave optics hold) and the medium is Kolmogorov turbulent \citep{Ocker2025CGMScattering}. Common sites of scattering include the Milky Way and host ISM, manifesting as interstellar scintillation. Multi-screen scintillation measurements suggest that the ubiquitous scattering tails observed in FRBs may originate near the source, potentially in the circumburst environment \cite{Ocker2022, Sammons2023}. \updated{Scattering from the circumgalactic medium of intervening halos has yet to be definitively detected (see \citet{Faber+2024} for a tentative detection), but scattering from the CGM could shed light on sub-grid physics in the CGM \citep[see][]{Caleb01.2026.SKA}.}

\subsection{Rotation Measure} \label{sec:rm}

The linearly polarised synchrotron emission from FRBs undergoes Faraday rotation due to free electrons and magnetic fields parallel to the line-of-sight, $\Blos$. Faraday rotation leads to the rotation of the polarisation plane as a function of the observing wavelength, $\lambda$, as $\Delta \psi = \RM\,\lambda^2$, where $\Delta \psi$ is the change in the polarisation angle between the source and observer and $\RM$ is the rotation measure, given by
% \begin{align}\label{eq:rm}
% \frac{\RM}{\rad\,\m^{-2}} = 0.812\,\int_0^{\dis/\pc} \frac{\ne}{\cm^{-3}} \frac{\Blos}{\muG} \dd \left(\frac{l}{\pc}\right).
% \end{align}
\begin{align}\label{eq:rm}
\RM = \frac{1}{2\pi}\frac{e^3}{m_e^2c^4}\,\int_0^{\dis} \ne \,\Blos \,\dd l\,,
\end{align}
where $e$ and $m_e$ are the electron charge and mass, respectively.
Like \Cref{eq:parts}, the $\RM$ of an FRB can also be expressed as a sum of different contributions,
\begin{align} \label{eq:rmparts}
    \RM_{\rm obs} (\vec{x}, z) = \RM_{\rm host}(z) + \RM_{\rm cosmic}(\vec{x}, z) + \RM_{\rm MW, ISM} (\vec{x}) + \RM_{\rm MW, CGM} (\vec{x}) + \RM_{\rm noise}(\vec{x}),
\end{align}
where $\RM_{\rm host}$ is the host contribution (including both the local environment and the host galaxy), $\RM_{\rm cosmic}$ is the contribution to magnetised plasma in the intergalactic medium and the large-scale structure, $\RM_{\rm MW, ISM}$ is from the Galactic ISM, $\RM_{\rm MW, CGM}$ is from the Milky Way's CGM, and $\RM_{\rm noise}$ quantifies the level of the fluctuations due to instrumental effects and analysis methods. Knowing $\ne$ from $\DM$ contributions, $\RM$ could be used to infer the properties of galactic and intergalactic magnetic fields. 

\section{Expected data with SKA} 
\label{sec:expected data}
In this section, we provide forecasts for the number of FRBs to be detected with SKA-mid and SKA-low for both AA$^*$ and AA4. The numbers presented here serve as the basis for the cosmological forecasts in the FRB chapters.

\begin{figure}
    \centering
    \includegraphics[width=0.49\linewidth]{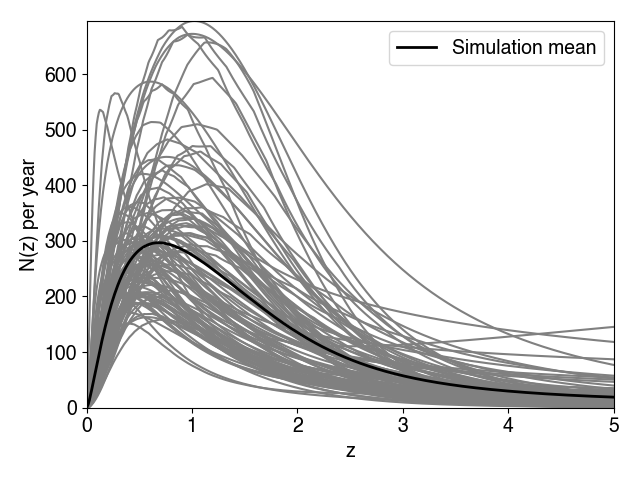} \includegraphics[width=0.49\linewidth]{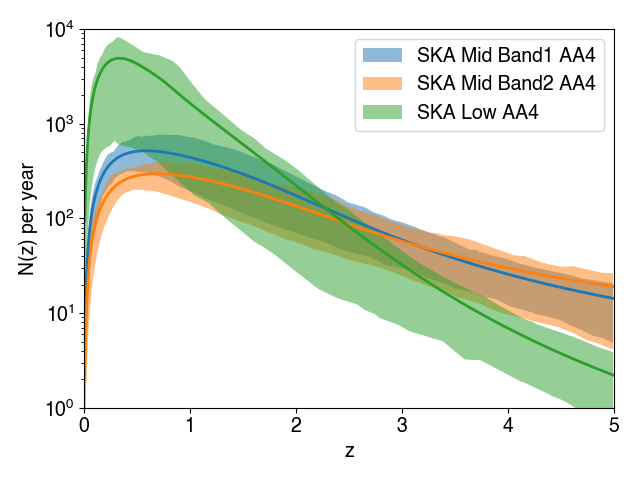}
    \caption{Predicted redshift distribution of FRBs detected by SKA. Left: predictions for band 2 of SKA Mid in AA4 configuration. The grey lines represent 100 instances of parameter sets fit to current FRB data; the black line represents the mean of these predictions. Right: predictions for different SKA observations, showing the means (lines), and 16--84\% quantiles (shaded). Normalisation is such that the total rate in a full year's worth of observation time is given by $\int N(z) \,\mathrm{d}z$.}
    \label{fig:Mid2AA4pz}
\end{figure}

\begingroup
\renewcommand*{\arraystretch}{1.2}
\begin{table*}[b]
    \centering
    \small
    \caption{Parameters used to model the FRB detection rate of SKA telescopes: mean frequency $\bar{\nu}$, bandwidth $\Delta \nu$, channel width $\delta \nu$, time resolution $t_{\rm res}$, field of view (FOV), detection thresholds $F_{\rm th}$ for AA4 and AA$^*$.}
    \begin{tabular}{l ccccccc}
    \hline
    Instrument & $\bar{\nu}$ & $\Delta \nu$ & $\delta \nu$ & $t_{\rm res}$ & FOV & $F_{\rm th}^{\rm AA4}$ & $F_{\rm th}^{\rm AA^*}$ \\
    & MHz & MHz & MHz & ms & deg$^2$ & Jy ms & Jy ms  \\
    \hline
      Mid Band 2 & 1400 & 300 & 0.1075 & 0.069 & 0.58 & 0.059 & 0.116  \\
      Mid Band 1  & 865 & 300 & 0.1075 & 0.069 & 1.79 & 0.091 & 0.167 \\
      Low     & 190  & 120 & 0.0145 & 0.069 & 5.12 & 0.143 & 0.164 \\
      \hline
    \end{tabular}
    \label{tab:instrument_properties}
\end{table*}
\endgroup

\subsection{Simulating SKA detection of FRBs}
Calculations of the number of FRBs detected by SKA telescopes, and their redshift--dispersion measure distribution, require three ingredients: properties of the detecting instrument, a treatment of the intrinsic FRB population, and a model for the cosmological distribution of ionised gas. Relevant properties of the former are given in Table~\ref{tab:instrument_properties}, corresponding to tied beams from the central stations being used to cover the FOV out to the half-power point. We do not include a detailed model of the tied-beam shape and assume a Gaussian primary beam. The FRB population is modelled with a Schechter luminosity function characterised by power-law slope $\gamma$ and turnover energy $E_{\rm max}$, a spectral index $\alpha$, and source evolution scaling with the star-formation rate modelled by \citet{2014ARA&A..52..415M} as SFR$(z)^n_{\rm sfr}$. Secondary parameters include the distribution of intrinsic FRB widths and scattering, which are taken from the fits of \citet{chime_2021_first}  ---- no redshift dependence of this has yet been included, since the fit was to observed (and not rest-frame) properties; however, scattering is assumed to scale as $\nu^{-4}$, and will be much more critical for SKA-Low. These parameters have been estimated based on FRB data from ASKAP, Parkes, FAST, and DSA by \citet{2025PASA...42...17H} \updated{using the \texttt{z-DM} code\footnote{\url{https://github.com/FRBs/zdm}}}. However, there are considerable uncertainties, so we iterate over 100 possible parameter sets chosen from the MCMC simulation of that work \updated{to illustrate the impact of these uncertainties on the forecasts}. For the cosmological DM budget, we assume a Milky Way contribution from both the ISM and halo of 80\,pc\,cm$^{-3}$, a log-normal distribution of FRB host dispersion measures with parameters encoded in the MCMC data of \citet{2025PASA...42...17H}, and cosmological contributions with mean, and fluctuations about the mean, described by \citet{macquart_census_2020}.

These three sets of ingredients are combined in the simulation {\texttt{$z$-DM}}, which produces estimates for the FRB distribution in terms of redshift, extragalactic dispersion measure, and FRB luminosity. Figure~\ref{fig:Mid2AA4pz} illustrates the expected redshift distribution in the case of SKA Mid band 2 in AA4 configuration --- the variance due to parameter uncertainties is approximately a factor of 2. However, this variance increases significantly for SKA-low, since much less is known about the behaviour of FRBs at low frequencies. Table~\ref{tab:numbers} lists the estimated annual rates. 

For these estimates it was implicitly assumed that an FRB survey is operating commenusally with all SKA observations. In practice, the actual number of FRBs detected will depend on the relative fraction of the time observed in the different frequency bands of SKA-mid, as well as the sky position.
The sample of FRBs predicted here are unique sources; dedicated follow-up of active repeaters may yield significantly higher numbers, which is not included in these estimates. Also, given the extremely uncertain detection rates at higher frequencies, we do not simulate estimates for SKA-mid bands 5a and 5b. \updated{Finally, these forecasts predict FRBs with DM $\lesssim$ 5000~pc~cm$^{-3}$ will be discovered with SKA-mid, which is larger than the currently planned maximum searched DM = 3000~pc~cm$^{-3}$. Given that the specifics of the SKA search pipeline are still in-progress and searches at high DMs have low computational cost, we make no cut in FRB DM in the following analyses.}

\subsection{Host visibility}

\begingroup
\renewcommand*{\arraystretch}{1.2}
\begin{figure}
    \centering
    \includegraphics[width=0.7\linewidth]{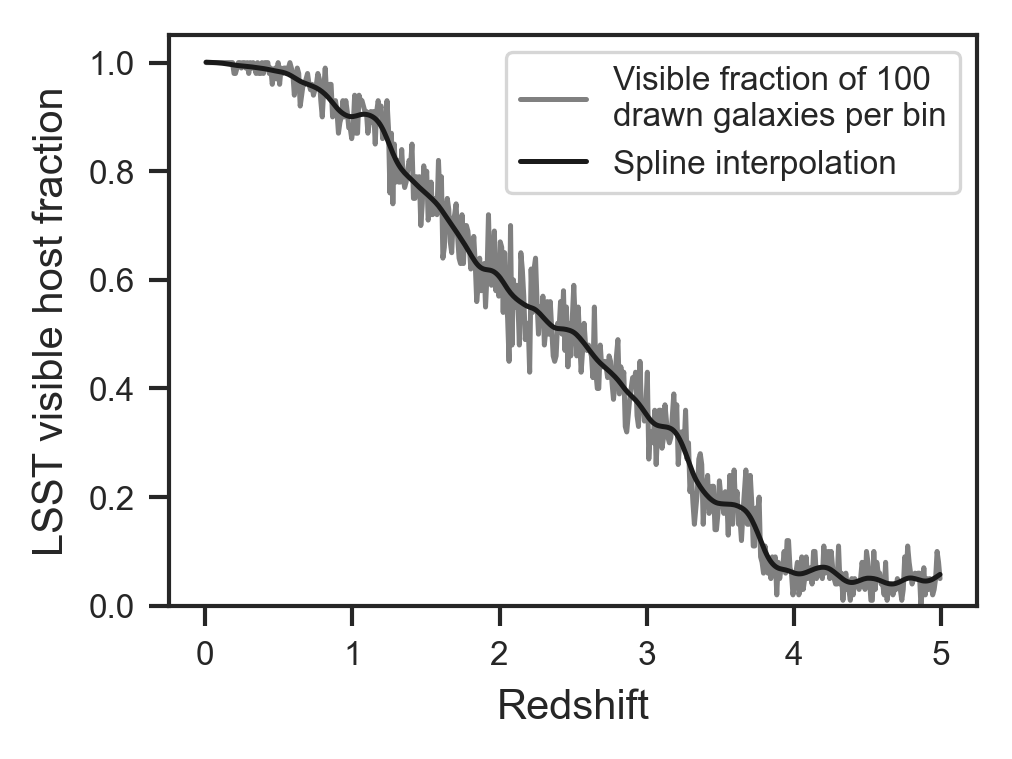}
    \caption{Fraction of FRB host galaxies that will be detected in one or more bands of the Rubin Observatory.}
    \label{fig:LSST}
\end{figure}
\begin{table}[]
    \centering
	\caption{Forecasted numbers of FRBs in one year of SKA on-sky observing and the fraction visible in the LSST. Reported numbers are the median, and the 16\% and 84\% quantiles.}
	\label{tab:numbers}
    \begin{tabular}{clccc}
        \hline
        Stage & Band & Nr.\ FRBs & Nr.\ visible hosts & Host fraction\\
        \hline
        \multirow{3}{*}{\rotatebox[origin=c]{90}{AA$^*$}} & Mid Band2 & $280_{-110}^{+120}$ & $220_{-80}^{+100}$ & $0.79_{-0.05}^{+0.05}$\\
         & Mid Band1 & $390_{-140}^{+270}$ & $320_{-110}^{+210}$ & $0.83_{-0.04}^{+0.04}$\\
         & Low & $1320_{-750}^{+5020}$ & $1230_{-700}^{+4800}$ & $0.94_{-0.02}^{+0.02}$\\
        \multirow{3}{*}{\rotatebox[origin=c]{90}{AA4}} & Mid Band2 & $550_{-210}^{+270}$ & $420_{-160}^{+200}$ & $0.75_{-0.05}^{+0.05}$\\
         & Mid Band1 & $790_{-290}^{+530}$ & $620_{-220}^{+440}$ & $0.79_{-0.04}^{+0.05}$\\
         & Low & $1490_{-860}^{+5700}$ & $1380_{-790}^{+5420}$ & $0.93_{-0.03}^{+0.02}$\\
        \hline
    \end{tabular}
\end{table}
\endgroup

The expected rate of FRB detections allows only for dedicated optical follow-up of a small fraction of the detections.
Therefore, host identifications have to rely primarily on existing catalogues.
Luckily, the SKA will show great synergy with the Rubin Observatory Legacy Survey of Space and Time \citep[LSST,][]{LSST}.

We estimate the fraction of hosts that will be visible over the full 10 years of LSST, following \citet{Jahns2023}.
We draw galaxies from their catalogue of galaxies simulated with the semi-analytic model of galaxy formation GALFORM \citep{Cole2000, Baugh2019}.
For each of the 500 redshift bins used in the FRB forecast, we draw 100 galaxies weighted by their star formation rate.
\updated{The simulated data include the observer frame brightness in each of the six Rubin Observatory bands.
For a galaxy to be visible, we require the brightness in at least one of the bands to exceed the target coadded 5-sigma depth after 10 years \citep{Bianco2022}.}
To apply the fractions to the previously simulated FRB numbers, we multiply each bin's observed fraction by its simulated FRB count and sum the products.

\Cref{fig:LSST} shows the redshift-dependent fraction.
The resulting numbers of observed FRB host galaxies are presented in Table~\ref{tab:numbers}.
The fraction is promising, spanning 0.75-0.94, although it becomes small above $z\sim3$.
It will likely be challenging to associate many galaxies with certainty above $z\sim2$ because of the high probability of an unseen host.

\begin{figure}
    \centering
    \includegraphics[width=0.7\linewidth]{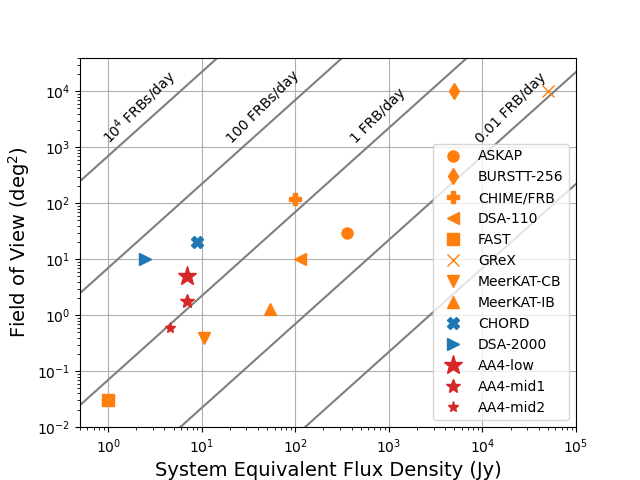}
    \caption{FRB detection rate parameter space plot. Current FRB surveys are shown in orange, upcoming surveys in blue, and the three SKA AA4 configurations and frequency bands discussed in the text in red. Based on the figure from \citet{BURSTT}.
    %\daniele{Maybe it could be clearer to put FRBs/day on the vertical axis, and FoV (or SEFD) slanted.} \laura{I'm not sure why changing the axes would be clearer.}
    }
    \label{fig:telescopes}
\end{figure}

\subsection{FRB Survey Landscape}
%\daniele{This could also be moved to the Intro, after current instruments.}
Advances in observational capabilities happen rapidly in the FRB field, and here we briefly discuss the future FRB survey landscape. Currently, two new telescopes with FRB science as one of the primary drivers are under construction. The Canadian Hydrogen Observatory and Radio-Transient Detector (CHORD) is being built in North America with the main telescope at the same site as CHIME/FRB and two outrigger stations with $>$1000~km baselines providing precise localisations \citep{CHORD}. Its sensitivity is comparable to that of SKA-AA4 configurations, but it has an instantaneous field of view that is a factor of $\sim$10 times larger (see Figure~\ref{fig:telescopes}). As a result, its detection rate is an order of magnitude larger than SKA-mid. CHORD will be equipped with receivers with a continuous bandwidth from 300--1500~MHz, spanning most of the gap between SKA-low and SKA-mid band1, and extending into most of SKA-mid band 2. 

The Deep Synoptic Array (DSA)-2000 is an array of small dishes to be constructed in Nevada, USA \citep{DSA2000}. It has high sensitivity and a moderate field of view, yielding a detection rate almost two orders of magnitude higher than that of SKA-mid AA4 configurations (Figure~\ref{fig:telescopes}). The operating range of the receivers is 0.7$-$2~GHz, which overlaps entirely with both SKA-mid bands and goes a few hundred MHz beyond in both directions. 

Both CHORD and DSA-2000 are scheduled to begin operation around 2027. At the time of writing, the start of coordinated or long-term projects with the AA$^*$ configuration is expected in 2032 and 2034 for SKA-low and SKA-mid, respectively. Moreover, CHORD and DSA-2000 expect to detect $\sim$10,000 FRBs per year and have instantaneous bandwidths larger than those of SKA-mid band 1 and band 2 combined. Therefore, by the time AA$^*$ comes online, CHORD and DSA-2000 will likely have been operating for several years and have potentially discovered $10^5$ FRBs.  

The SKA, CHORD, and DSA-2000 are all narrow but deep surveys. In recent years, interest in wide but shallow surveys has been growing. These telescopes are aperture arrays of low-cost antennas that generally operate just below $\sim$1~GHz. BURSTT-256 is an example that is currently operational in Taiwan \citep{BURSTT}, but several additional arrays are under development worldwide. In the near future, these telescopes cannot compete in the absolute number of discoveries, but they can fill a role, for example, in discovering a population of local Universe sources, including Galactic FRBs.

\section{Magnetic Fields in the High-Redshift Universe} \label{sec:mag}

Magnetic fields are present and important in a variety of astrophysical systems but are often challenging to measure observationally beyond the local Universe \citep{KleinF2015, ShukurovS2021}. FRBs could be very useful for probing magnetic fields in the early Universe, especially in high-redshift galaxies and in the intergalactic medium (IGM).

The $\DM_{\rm cosmic}$ and $\RM_{\rm cosmic}$ terms in \Cref{eq:parts} and \Cref{eq:rmparts}, respectively, include contributions from both intervening galaxies and the IGM (including cosmic filaments and voids). These terms can be further decomposed as
\begin{align}
\DM_{\rm cosmic} & = \sum_{i=1}^{\Ningal} \dfrac{\DM_i(\vec{x})}{(1 + z_{\ingal,i})} +  \DM_{\rm IGM} (\vec{x}, z), \label{eq:dmcos} \\
\RM_{\rm cosmic} & = \sum_{i=1}^{\Ningal} \dfrac{\RM_i(\vec{x})}{(1 + z_{\ingal,i})^2} + \RM_{\rm IGM} (\vec{x}, z), \label{eq:rmcos}
\end{align}
where $\DM_i$ and $\RM_i$ denote the contributions from the $i$th intervening galaxy at redshift $z_{\ingal,i}$, with $\Ningal$ being the total number of intervening galaxies along the line of sight. The terms $\DM_{\rm IGM}$ and $\RM_{\rm IGM}$ represent the contributions from the IGM. 

Below, in \Sec{sec:mag_gal} and \Sec{sec:mag_igm}, we discuss the significance of magnetic fields in high-redshift galaxies and in the IGM, along with possible approaches to study them with SKA.
 
\subsection{Magnetic fields in high-redshift intervening galaxies} \label{sec:mag_gal}
Magnetic fields in the early Universe are observed to be orders of magnitude weaker \citep[$\approx 10^{-10}\,\muG$, see][]{KulsrudEA1997, Subramanian2016, SetaF2020} than those in present-day galaxies \citep[$\approx 10^{1}\,\muG$, see][]{Beck2015, Haverkorn2015, SetaMCG2025}. This amplification is believed to result from a dynamo process that converts the kinetic energy of turbulence into magnetic energy \citep{RuzmaikinEA1988, BrandenburgS2005, Rincon2019, ShukurovS2021}. Such amplification is seen in a variety of magnetohydrodynamic simulations \citep{SchekochihinEA2004, HaugenEA2004, GentEA2013, SetaEA2020, SetaF2021, GentEA2023} but convincing observational evidence is still lacking. There have been some efforts in this direction using $\RM$ from radio galaxies, including both statistical studies \citep{BernetEA2008, FarnesEA2014, MalikEA2020, ShahS2021, AmaralEA2021} and individual detections in gravitationally lensed systems \citep{MaoEA2017, KovacsEA2025}. However, the statistical samples are highly inhomogeneous, and the lensed systems are biased toward more massive galaxies. Most importantly, both approaches require assumptions about the thermal electron density in order to extract magnetic field information from the observed $\RM$, and these properties are highly uncertain. With $\DM$s from FRBs, the magnetic field can be constrained more accurately, and with dense redshift sampling, its properties as a function of redshift can be mapped. Such an in-depth study would help constrain dynamo theories and improve galaxy evolution simulations.

\begin{figure*}
    \includegraphics[width=0.5\columnwidth]{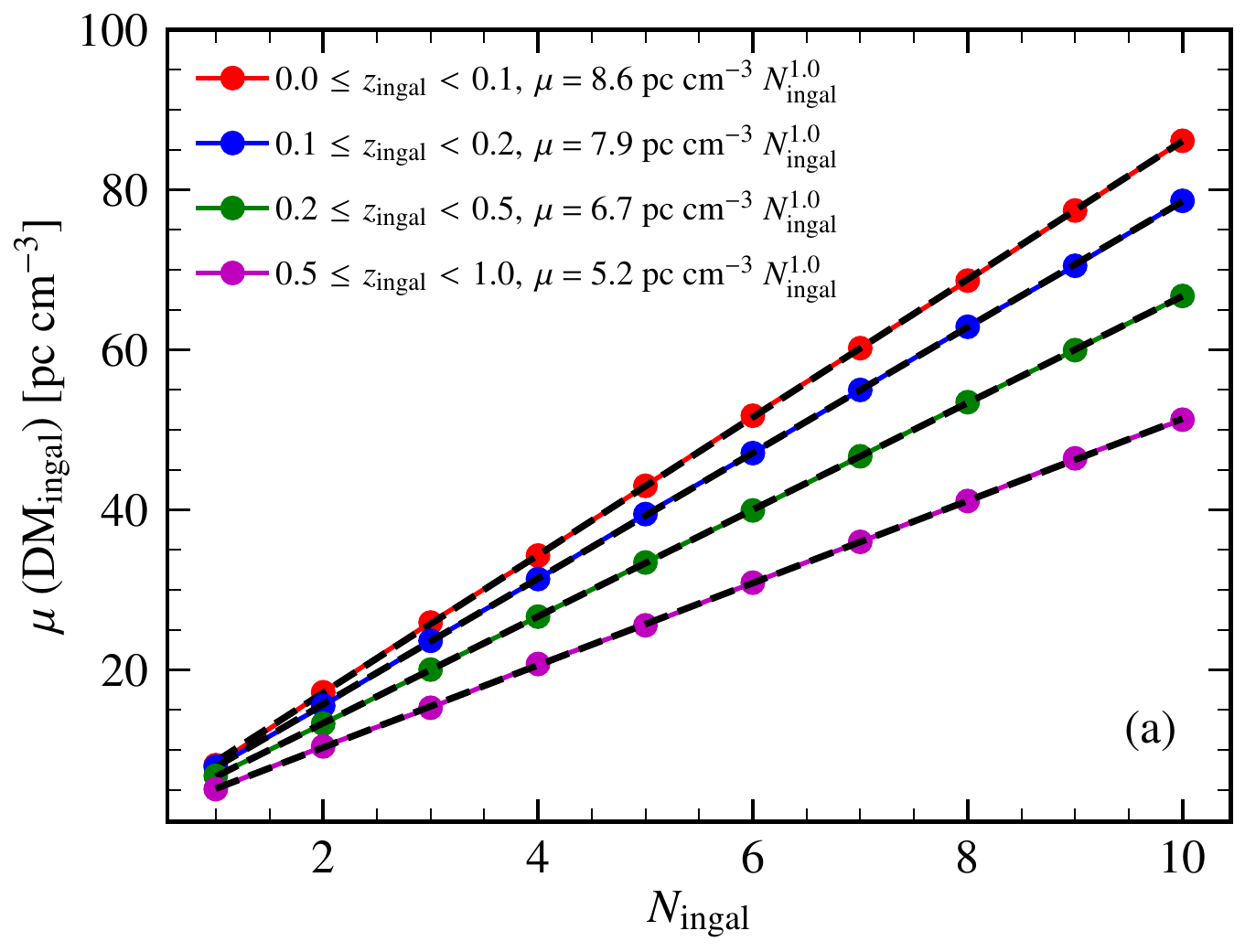} \hspace{0.05cm}
    \includegraphics[width=0.5\columnwidth]{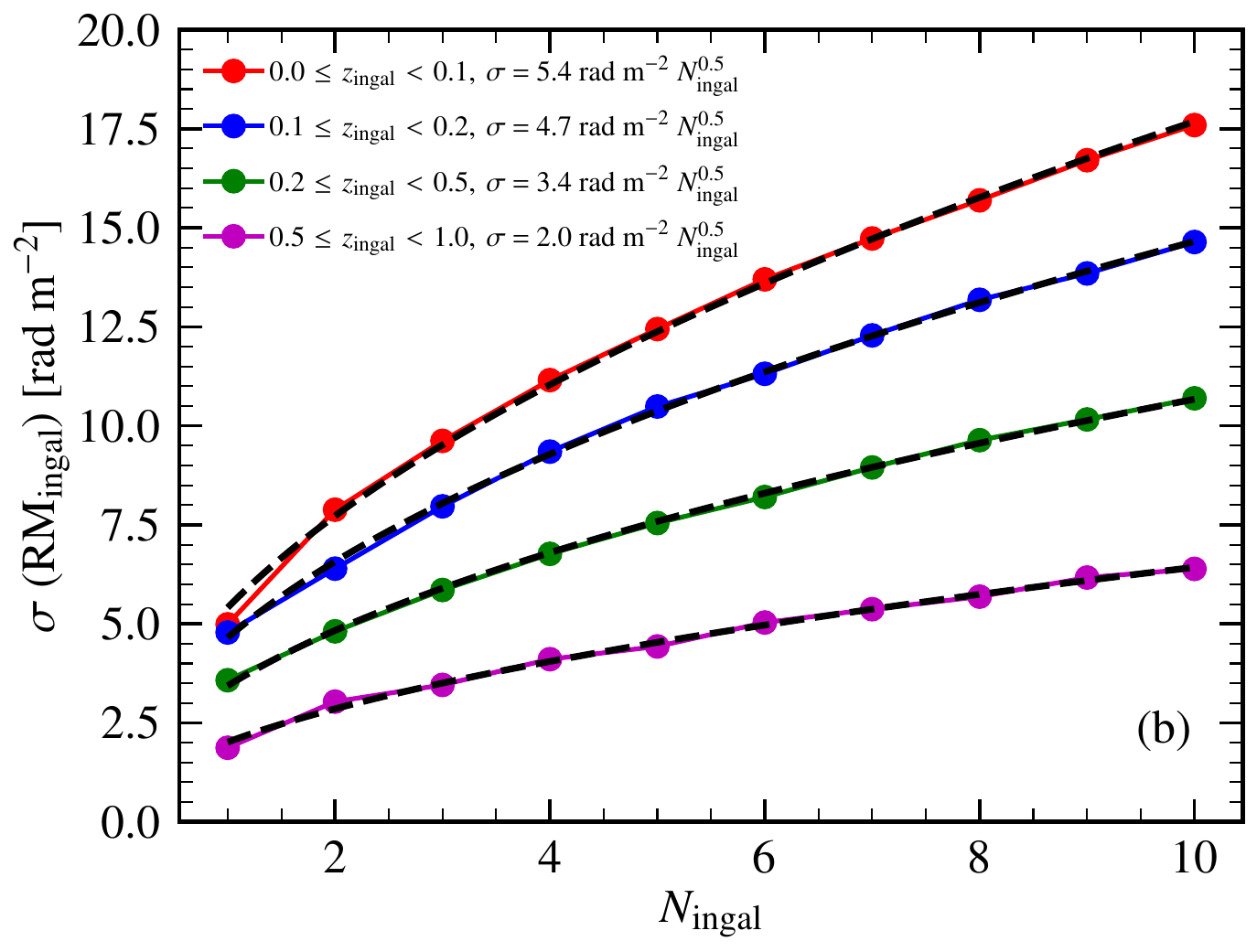}
    \caption{Mean and standard deviation of the dispersion and rotation measures from intervening galaxies, $\mu~(\DM_{\ingal})$ (a) and $\sigma~(\RM_{\ingal})$ (b), for galaxies at different redshifts (see legend). These values are based on simple analytic models for the thermal electron density and magnetic fields (see text). Across all redshift bins, $\mu~(\DM_{\ingal}) \propto \Ningal^{1}$ and $\sigma~(\RM_{\ingal}) \propto \Ningal^{0.5}$.}
    \label{fig:mag_ingal}
\end{figure*}

Several previous studies have used cosmological simulations to investigate the $\RM$ and $\DM$ contributions from each component in \Cref{eq:parts} and \Cref{eq:rmparts} \citep{AkahoriEA2016, HacksteinEA2019, HacksteinEA2020, KovacsEA2024}, though some did not further decompose $\DM_{\rm cosmic}$ and $\RM_{\rm cosmic}$ as in \Cref{eq:dmcos} and \Cref{eq:rmcos}. \updated{A recent investigation with the FLIMFLAM survey does indicate that the contribution of ${\rm RM}_{\rm ingal}$ to the observed FRB RMs is non-negligible \citep{KhrykinEA206}.} We expect the total contributions of $\DM_{\ingal}$ and $\RM_{\ingal}$ to correlate with the number of intervening galaxies, $\Ningal$. Once $\DM_{\rm cosmic}$ and $\RM_{\rm cosmic}$ are determined (already a challenge given the uncertainties in other terms in \Cref{eq:parts} and \Cref{eq:rmparts}), a fraction of them should correlate with $\Ningal$ \citep[such a correlation has been hinted at in $\DM$ observations, see][]{HussainiEA2025}. Here, we estimate the dependence of $\DM_{\ingal}$ and $\RM_{\ingal}$ on $\Ningal$. Quantifying this dependence would allow us to subtract the contributions of intervening galaxies and thereby isolate the IGM terms, providing a way to constrain magnetic fields in the IGM (see \Sec{sec:mag_igm} for further discussion).

It is also expected that FRB sightlines intersect primarily the CGM of intervening galaxies rather than their ISM, with typical impact parameters in the range $1$--$100\,\kpc$. The connection between ISM and CGM dynamics, and therefore their magnetic fields, is an area of active research \citep{PakmorEA2020, vanEA2021, ShahEA2025}. Consequently, we adopt simple models for the thermal electron density and magnetic fields in the CGM with a typical size of $200\,\kpc$ to estimate how the total $\DM_{\ingal}$ and $\RM_{\ingal}$ depend on $\Ningal$. For the thermal electron density, we assume a spherically symmetric beta model \citep[e.g.~see][]{MillerB2013}, $\ne(r) =  5 \times 10^{-4}\,\cm^{-3} \, \bigl(1 + (r/10\,\kpc)^2\bigr)^{-3/4}$. For the magnetic field magnitude, we use an analytic model guided by radio polarisation observations of nearby galaxies at smaller radii and by cosmological simulations at larger radii \citep[e.g.~Fig.~8 in][]{ShahS2021}. Further, for the magnetic field scale, based on the large-scale field in nearby galaxies \citep{Beck2015, SetaF2024}, we assume that the magnetic field may change sign (due to a random process) on a $1\,\kpc$ scale along the path length.

\Cref{fig:mag_ingal} shows the mean of $\DM_{\ingal}$ (a) and standard deviation of $\RM_{\ingal}$ (b) as a function of $\Ningal$ for various redshift bins of the intervening systems. We find that $\mu~(\DM_{\ingal}) \propto \Ningal^{1}$ and $\sigma~(\RM_{\ingal}) \propto \Ningal^{0.5}$ with the magnitude for both decreasing with increasing redshift. Searching for such trends in the FRB sample with known intervening galaxies along the line of sight will help isolate the contribution from intervening galaxies. We note that these magnitudes might also depend on the assumed thermal electron density and magnetic fields but the overall trend should be consistent with more complex models, such as cosmological simulations.

\begin{table}[]
    \centering
	\caption{$\RM$ uncertainties given the band for AA4 stage of SKA.}
	\label{tab:rms}
    \begin{tabular}{lcc}
        \hline
        Band & Frequency Range [MHz] & RM uncertainty [$\rad\,\m^{-2}$]\\
        \hline
        SKA-Low & $50$--$350$ & $0.008$\\
        SKA-Mid, Band 1 & $350$--$1050$ & $0.4$\\
        SKA-Mid, Band 2 & $950$--$1760$ & $4$\\
        \hline
    \end{tabular}
\end{table}

From the order-of-magnitude values in \Cref{fig:mag_ingal}, $\DM$ variations should be detectable given the $\DM$ uncertainty of $\lesssim 1\,\pc\,\cm^{-3}$. In contrast, the minimum reliably measurable $\RM$ depends on the frequency coverage. Estimating these limits \citep[see][and assuming a signal-to-noise ratio of 6]{BrentjensB2005} for the AA4 configuration of the SKA (see \Tab{tab:rms}) shows that detecting the excess $\RM$ from intervening galaxies is feasible.

\subsection{Magnetic fields in the intergalactic medium} \label{sec:mag_igm}

The origin of magnetic fields in the Universe remains an open question. In particular, the physical origin of the seed fields, which are later amplified by dynamo mechanisms in turbulent astrophysical environments (as discussed in \Sec{sec:mag_gal}) is unknown. These seed fields could be either primordial or astrophysical in origin \citep{WidrowEA2012, DurrerN2013, Subramanian2016}. Primordial fields may have been generated in the early Universe during inflation or phase transitions, while astrophysical seed fields could arise from the Biermann battery mechanism, outflows from the first stars and AGNs, or plasma instabilities. Thus, it is important to observationally probe magnetic fields in the IGM, especially in cosmic filaments and voids, where FRBs could be of use.

FRB sightlines without detectable intervening galaxies or clusters and/or cases where their contributions can be statistically accounted for in \Cref{eq:dmcos} and \Cref{eq:rmcos} are useful for probing magnetic fields in the IGM. This approach complements other FRB-based methods, such as scattering and scintillation, for studying IGM magnetism \citep[][]{RaviEA2016, PadmanabhanEA2023}. The IGM contribution, however, is expected to be small ($\lesssim 1\,\rad\,\m^{-2}$), so even with the large number of FRBs expected from the SKA, a statistical analysis with multiple such lines of sight will likely be required.

\section{Expansion of the Universe} \label{sec:expansion}
FRBs can be used as distance estimator and therefore can be used to test the expansion rate of the Universe. There are different methods to tackle this. Here we discuss two options: $(i)$ using the Macquart relation and $(ii)$ exploiting strongly lensed FRBs. \updated{A more quantitative presentation of the constraints on the Hubble constant based on the simulations presented here is given in the companion chapter \citet{Caleb01.2026.SKA}.}

\subsection{The Macquart relation}
Since host-identified FRBs provide both a redshift estimate and an independent distance measure (their DM), the DM-$z$ relation offers a direct test of the background cosmology (see Eq. \ref{eq:DM_LSS_v2}). In several papers, the possibility of using FRBs to measure the expansion history of the Universe has been explored in detail \citep[e.g.][]{walters_future_2018,walters_probing_2019,2022MNRAS.511..662H}. These studies show that, from a large sample of FRBs, it becomes possible to constrain cosmological parameters, such as the Hubble constant ($H_0$), matter density ($\Omega_\mathrm{m}$), and potentially even properties of dark energy, while marginalising over both the host and Milky Way contribution. This shows the FRB's potential to complement traditional methods such as supernovae and baryon acoustic oscillations. 
With the SKA, the number of available FRBs is expected to increase substantially. For the SKA numbers presented in \Cref{sec:expected data}, we can anticipate having independent constraints at the level of $1\,\%$ on those key cosmological parameters.
As discussed in the introduction, other experiments will yield similar FRB detection rates, and achieving a further order-of-magnitude increase in sensitivity would require detecting $10^2$ more FRBs than current experiments, which is unlikely in the near term. Nevertheless, the SKA will contribute two critical complementary advantages. First, it will detect FRBs at high redshifts, thereby allowing us to probe the Universe at earlier epochs. Second, the SKA is uniquely positioned as the only telescope covering extensive areas of the southern hemisphere at these redshifts. This broad sky coverage will ensure that the sightlines to potential FRB detections are less correlated than they would be if all observations were restricted to the northern hemisphere. As a result, the SKA's observations will offer greater statistical independence and provide much-needed robustness against systematic effects as the FRB sample size grows. In conclusion, while the SKA may not deliver fundamentally new constraints on the expansion history itself, it will undoubtedly serve as a vital robustness check in an era where systematic effects increasingly dominate cosmological measurements.

\subsection{Lensed FRBs}
The delay between multiple images of a strongly lensed source whose luminosity varies with time can be used to estimate the Hubble constant (\hubble) and constrain additional cosmological parameters, a technique known as time-delay cosmography \citep{Birrer2024}.
Variable background sources used in the past include AGNs and, more recently, supernovae \citep{Grillo2024}.
With their short duration, FRBs represent ideal sources for time-delay cosmography \citep{Li2018}.
Using FRBs, it could even become feasible to measure cosmological parameters without modeling the mass profile of the lens \citep{Wucknitz2021}, removing the largest source of uncertainty in current results \citep{Millon2020}.
However, finding FRBs strongly lensed by galaxies (L-FRBs) is challenging due to the rarity of the phenomenon. In fact, their lensing optical depth is of the order of $10^{-3}$ -- $10^{-4}$, depending on the source distance \citep{Yue2022}. 
Thus, finding L-FRBs requires sensitive telescopes capable of discovering thousands of FRBs per year. 
This is in the range of upcoming facilities such as CHORD and DSA-2000.
The discovery rates presented in Table~\ref{tab:instrument_properties} imply that SKA could also contribute to this endeavor, detecting an order of one L-FRB per year across all observing bands.

Although SKA-Low is expected to find many more FRBs, the effect of pulse broadening due to multipath propagation in the lens galaxy could negatively impact the discovery rate of L-FRBs at lower frequencies \citep{Ocker2022b}. 
The extent of this effect is difficult to estimate due to the unknown degree of turbulence in the plasma of lens galaxies. 
However, since most lenses are elliptical and impact parameters in the lens plane are usually large, pulse broadening is probably low for most L-FRBs.
Additionally, it could be possible to increase the detection rates by lowering the threshold of FRB searches in the direction of sightlines with high lensing optical depth, such as clusters \citep{Sammons2025}.

A second challenge in using L-FRBs for time-delay cosmography comes from the necessity of detecting multiple images of the same source \citep{Connor2023}. 
Given typical delays of days to weeks and the limited fields of view of sensitive telescopes, the chance of detecting both images coincidentally is low. 
Thus, it would be necessary to identify L-FRBs detected in FRB searches, then predict the arrival window of subsequent images with a model of the lens, and finally use multiple telescopes worldwide to observe the second image. 
Given that the second image is usually weaker, sensitive telescopes would facilitate this endeavor, and SKA could play a role in this regard whenever its field of view overlaps with FRB hunting machines such as DSA-2000.

\section{Tests of dark matter} \label{sec:dark}

\subsection{Dark matter in host galaxies}
The host contribution to the DM essentially measures a combination of the gas densities and source distributions. It is, therefore, an excellent test bed for physical processes on the scales of galaxies and is sensitive to astrophysical processes. Physically, the statistical properties of the host contribution arise from the probability distribution function of FRBs in galaxies as well as the distribution of the hot gas. 
The exact mechanism responsible for FRBs is still under debate. In almost all cases, however, they are associated with stars and are hence chosen to follow the stellar density. 

 Ultra-light dark matter such as axion-like particles \citep[ALPs, see][for a review]{2016PhR...643....1M} form stable, localised field configurations. These configurations are known as solitons or axion mini-clusters. They may potentially alter the inner regions of density profiles observed in galactic halos. 
Physically, this effect can be interpreted as a quantum mechanical pressure. This arises due to the de Broglie wavelength of particles with very low mass. The wavelength can be on the scale of galaxies. 
As a result, the DM profile differs greatly from that of cold dark matter. These effects alter the underlying gas and stellar profiles in halos. The host contribution can therefore be used to measure the effect of ultra-light dark matter particles via their effect on the corresponding density profiles.

\begin{figure}
    \centering
    \includegraphics[width=0.7\linewidth]{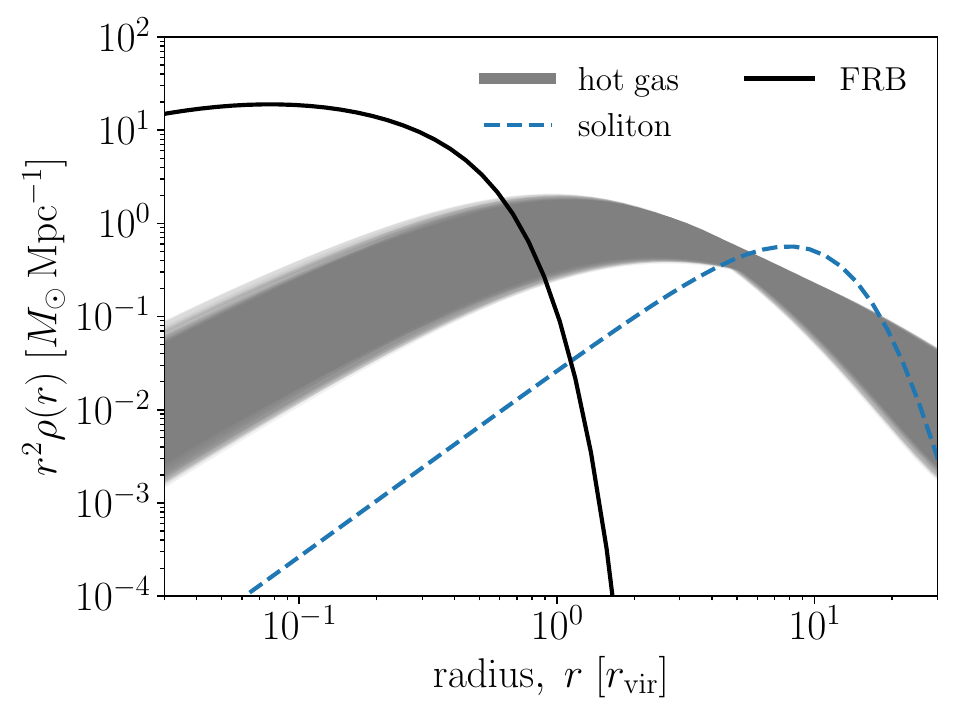}
    \caption{Normalised density profiles in an Milky Way-sized halo. Shown are stars (i.e. FRBs, solid black), the hot gas (free electrons, grey shaded area). The blue dashed profile shows a solitonic core of an axion-like particle of mass $m_\mathrm{ALP} = 10^{-23}\;\mathrm{eV}$.}
    \label{fig:soliton}
\end{figure}

\updated{Example profiles are shown in \Cref{fig:soliton} for a Milky Way-like galaxy. The black line shows the distribution of stars, which indicates where FRBs will statistically originate from in the galaxy. In dashed blue the solitonic core of an ALP is shown. The grey area indicates the gas profile with uncertainties derived from \citep{2025arXiv250707991K}. While there exist no quantitative forecasts of this effect so far, the presence of ALPs will alter both the hot gas distribution, as well as the stellar density profile, and hence change the contribution expected from the host galaxy. In particular, pinpointing the FRB's position within the host itself, which will be possible at low redshift thanks to the SKAO's excellent resolution, may make it possible to statistically disentangle the different density profiles and put lower limits on the ALP mass as a lower ALP mass increases the effect and size of the solitonic core. }

\subsection{Microlensing}
Microlensing by stellar-mass bodies produces multiple images of a background source. However, these micro-images are separated by orders of microseconds, too small to be resolved with current instruments. 
Thus, microlensing is at present only observed as a flux variation of the background source due to magnification.
With FRBs, it could be possible to resolve, for the first time, micro-images in the time domain \citep{Kader2022}.
In fact, the coherence of FRB signals is preserved by gravitational lensing, and thus, multiple micro-images can be identified in the autocorrelation of voltages measured by telescope antennas at the time of an FRB.

The detection of microlensing in the signal of FRBs has multiple applications, such as constraining the density of primordial black holes \citep{Leung2022} and of MACHOs \citep{Munoz2016}, identifying L-FRBs \citep{Sathyanathan2025}, assessing properties of lens galaxies \citep{Meena2025}, and even measuring cosmological parameters \citep{Tsai2024}.
The SKA sensitivity will allow us to observe a large volume of the Universe.
For example, the volume probed and its FRB discovery rate make SKA competitive in constraining the density of primordial black holes.

One potential show-stopper for gravitational microlensing is scattering. Identifying a gravitationally FRB relies on correlating the voltage signal between the two images, but if the images travel through different plasmas, multi-path propagation effects may alter their phases. Scattering can also affect magnification and time delays \citep{KB23}. To avoid these effects, FRB surveys in the higher frequency bands, including SKA-mid bands 5a/5b are particularly powerful. We also note that need for voltage data for this science goal requires storing baseband data around detected FRBs.

\section{Fundamental Physics} \label{sec:fundi}
Effects from quantum gravity or other fundamental principles are usually complicated to detect using terrestrial experiments. Highly energetic astrophysical transients, however, are an excellent testbed for these effects due to the large distances covered, and hence the cumulative impact of intrinsically minor effects is lifted above the noise.
The observed time delay, $\Delta t_\mathrm{obs}$, between frequency bands of an FRB comprises several contributions:
\begin{equation}
      \Delta t_\mathrm{obs} = \Delta t_\mathrm{int} + \Delta t_\mathrm{LIV} + \Delta t_\mathrm{m} + \Delta t_\mathrm{grav}.
\end{equation}
Here, $\Delta t_\mathrm{int}$ is an intrinsic delay, which includes the DM contribution $\Delta t_\mathrm{DM}$ and a possible source contribution $\Delta t_\mathrm{s}$. These delays are used in \Cref{eq:dm_definition} to infer the dispersion. However, the additional terms will lead to greater dispersion. The term $\Delta t_\mathrm{LIV}$ relates to Lorentz invariance violation (LIV), $\Delta t_\mathrm{m}$ could arise from dispersion if photons are massive, and $\Delta t_\mathrm{grav}$ is a (Shapiro) gravitational time delay. A lot of these tests can also be done with gamma-ray bursts. These are intrinsically more sensitive per object. However, FRBs are more abundant and measure these effects at a different energy scale.

\subsection{Lorentz invariance}
If Lorentz invariance is violated, the dispersion relation for photons can be expanded in the following model-independent form:
\begin{equation}
    E^2 \simeq p^2c^2\left[1-s_n\left(\frac{pc}{E_{\mathrm{LIV},n}}\right)^n\right],
\end{equation}
where $E_{\mathrm{LIV},n}$ is the energy scale where LIV occurs expanded in $n$ orders. $s_n$ simply controls the sign of the LIV and whether these photons travel faster or slower than $c$. Calculating the group velocity allows us to infer the additional time delay
\citep{2008JCAP...01..031J}:
\begin{equation}
    \Delta t_{\mathrm{LIV}}=s_{n}\frac{1+n}{2}\frac{E_{1}^{n}-E_{2}^{n}}{E_{\mathrm{LV},n}^{n}}\int_{0}^{z}\mathrm{d}z'\frac{(1+z')^{n}}{H(z')},
\end{equation}
here $E_{1}$ and $E_{2}$ are the upper and lower energies of the photons measured in the frequency band. Thus, decomposing the time delay into components with different frequency scalings allows separation of LIV terms, thereby constraining the amount of LIV at low photon energies.

\begin{figure}
    \centering
   \includegraphics[width=0.45\textwidth]{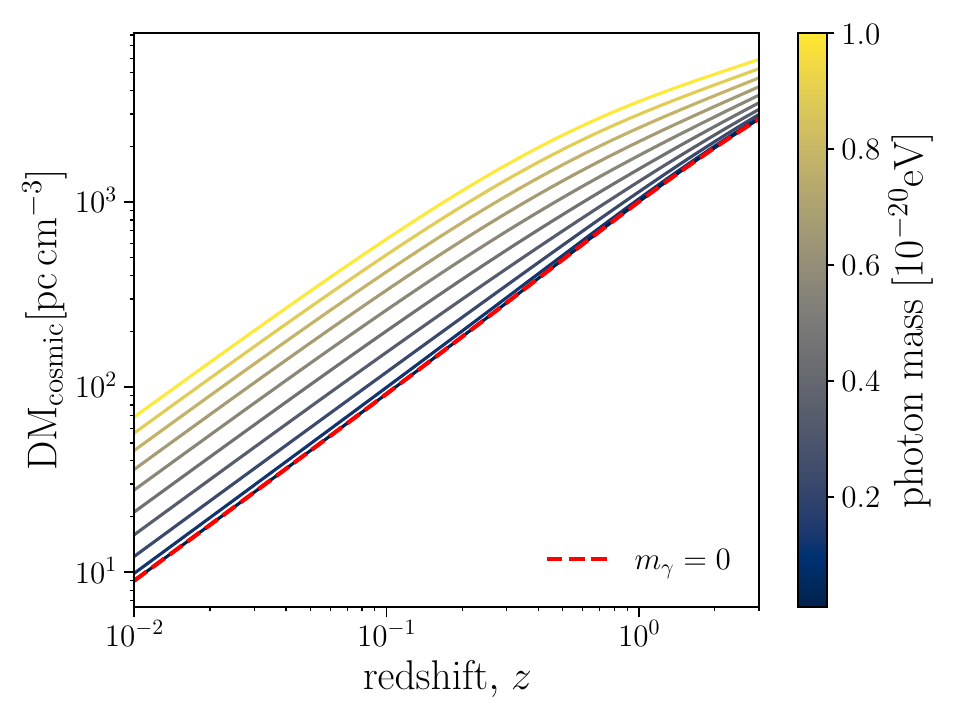}
   \includegraphics[width=0.45\textwidth]{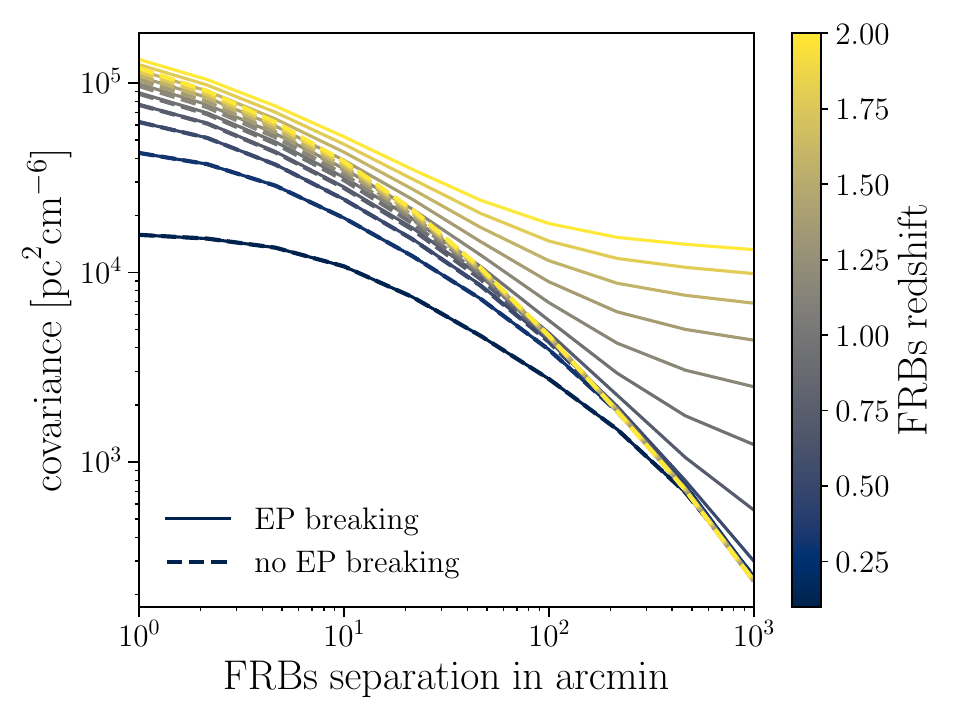}
    \caption{\textit{Left}: The effect of massive photons on the cosmological DM, \Cref{eq:DM_massive}. The dashed red line indicates the case where photons have vanishing mass, and the colourbar indicates the mass of the photons. \textit{Right}: The covariance between different sightlines of FRBs as a function of their pairwise angular separation and the redshift of that pair as a colour scale. EP breaking is indicated as solid lines and the standard covariance as dashed lines.}
    \label{fig:fundamental_physics_signal}
\end{figure}

\subsection{Massive photons}
$\Delta t_\mathrm{m}$ could arise from dispersion if photons are massive. This leads to a modified relativistic dispersion relation:
\begin{equation}
     E^2 = c^2 p^2 + m_\gamma^2 c^4,
\end{equation}
in turn resulting in an additional time delay \citep[see e.g.][]{2017PhRvD..95l3010S}
\begin{equation}
\label{eq:DM_massive}
     \Delta t_\mathrm{m}(z) = \left(\frac{m_\gamma c^2}{4 \pi \hbar} \right)^2 \left( \nu_1^{-2} - \nu_2^{-2} \right) \int_0^z  \frac{\mathrm{d} z'}{(1+z')^2 H(z')} ,
\end{equation}
where $\nu_1^{-2}$ and $\nu_2^{-2}$ are the frequencies at which the time delay is measured.  
Due to the frequency dependence, this delay will contribute to the overall measured dispersion. Both $\Delta t_\mathrm{LIV}$ and $\Delta t_\mathrm{m}$ cause non-zero effects on the cosmological background, acquiring additional DM. The left panel of \Cref{fig:fundamental_physics_signal} shows the expected signal from $m_\gamma \neq 0$, showing that massive photons lead to an additional delay similar to $\mathrm{DM}_\mathrm{cosmic}$ but with a different redshift scaling due to the changed dispersion relation.

\subsection{Equivalence principle}
Since FRBs probe cosmological potentials, the standard Shapiro delay equation cannot be used due to its explicit gauge dependence and divergence \citep[this is discussed in][]{2019PhRvD.100j4047M,2022MNRAS.512..285R,2023MNRAS.523.6264R}.
In a cosmological context, these issues can be mitigated by using a weakly perturbed Friedman-Robertson-Walker line element in the conformal Newtonian gauge within the parametrised post-Newtonian formalism framework:
\begin{equation}
    \label{eq:metric}
    \mathrm d s^2 = -\left(1+ \frac{2 \phi}{c^2} \right)  c^2 \mathrm{d} t^2 + a^2(t)  \left(1 - \frac{2 \gamma \phi}{c^2} \right)  \mathrm{d} \mathbf{x}^2\;,
\end{equation}
where $\phi$ is the gauge potential, $a$ the scale factor, and $\mathbf{x}$ the comoving coordinates. The time delay between photons of varying frequencies is given by 
\begin{equation}
    \Delta t_\mathrm{grav}(\hat{\boldsymbol{x}}) = \frac{\Delta\gamma}{c^3}\int^{\chi_\mathrm{s}}_{0} \mathrm{d}\chi a(\chi)  \phi \big( \hat{\boldsymbol{x}}\chi,a(\chi) \big).
\end{equation}
where $\chi$ represents the comoving distance at a background level. This formula does not diverge because it adheres to cosmological symmetry assumptions. Although it resembles the standard Shapiro delay equation, the perturbation $\phi$ acts as a random field with zero mean, allowing time delay to receive both positive and negative contributions along the line-of-sight \citep[see also][in the context of gamma ray bursts]{2021PhRvD.104j3516B}.

\begingroup
\renewcommand*{\arraystretch}{1.2}
\begin{table}[]
    \centering
	\caption{Expected uncertainties given the different stages and frequency bands of SKA for the mass of the photon, $m_\gamma$, and the equivalence principle breaking parameter, $\Delta\gamma$. Each combination of stage and band is given by to possible uncertainties, $\sigma_\mathrm{opt}$ and $\sigma_\mathrm{pess}$, corresponding to the best and the worst case from the simulated $N(z)$ in terms of number of FRBs (compare to \Cref{fig:Mid2AA4pz}). All numbers are for a five years of observations.} 
	\label{tab:fundamental_physics}
    \begin{tabular}{clcr}
            \hline
    & & \multicolumn{2}{c}{uncertainty as [$\sigma_\mathrm{opt}$,$\sigma_\mathrm{pess}$]}\\
        stage & band & $m_\gamma\times 10^{-23}$[eV] & $\Delta \gamma\times10^{-14}$\\
        \hline
        \multirow{3}{*}{\rotatebox[origin=c]{90}{AA$^*$}} & SKA-Low & $[0.28,6.37]$ & $[0.81, 14.3]$ \\
& SKA-Mid, Band 1 & $[1.26,4.63]$ & $[2.28, 9.82]$ \\
& SKA-Mid, Band 2 & $[1.51,4.62]$ & $[2.51, 10.7]$ \\
\hline
\multirow{3}{*}{\rotatebox[origin=c]{90}{AA$4$}} & SKA-Low & $[0.22,5.67]$ & $[0.63, 11.2]$ \\
& SKA-Mid, Band 1 & $[0.89,3.31]$ & $[1.83, 7.61]$ \\
& SKA-Mid, Band 2 & $[1.02,3.42]$ & $[2.78, 8.05]$ \\
        \hline
    \end{tabular}
\end{table}
\endgroup

Any extra delay in arrival time between pulse frequencies can indicate a violation of the equivalence principle (EP)\footnote{Our use of EP here refers to the weak equivalence principle.}. This extra dispersion is
\begin{equation}
\label{eq:dispersion_measure_observed}
    \mathrm{DM}_\mathrm{obs}(\hat{\boldsymbol{x}},z) \to \mathrm{DM}_\mathrm{obs}(\hat{\boldsymbol{x}},z) + \mathrm{DM}_\mathrm{grav}(\hat{\boldsymbol{x}},z).
\end{equation}
$\mathrm{DM}_\mathrm{grav}(\hat{\boldsymbol{x}},z)$ is the weak EP breaking term expressed as a DM along $\hat{\boldsymbol{x}}$ out to redshift $z$:
\begin{equation}
\label{eq:dispersion_measure_grav}
    \mathrm{DM}_\mathrm{grav}(\hat{\boldsymbol{x}},z) = \frac{\Delta\gamma}{\mathcal{K}c^3\left(\nu^{-2}_1-\nu^{-2}_2\right)}\int_0^{\chi(z)}\mathrm{d}\chi^\prime a(\chi^\prime)  \phi\big(\hat{\boldsymbol{x}}\chi^\prime,z(\chi^\prime)\big),
\end{equation}
where $\mathcal{K} = e^2/(2\uppi m_\mathrm{e} c)$.
Note that $\langle\mathrm{DM}_\mathrm{grav}(\hat{\boldsymbol{x}},z)\rangle = 0$, i.e., on average there is no contribution to the mean dispersion (i.e., to the Macquart relation). However, one can use the scatter in the DM as a measure of the EP.

\citet{2023MNRAS.523.6264R} calculated the covariance in the DM if the EP is broken. In \Cref{fig:fundamental_physics_signal} on the right, the covariance is shown as a function of the angular separation between two FRB sight-lines with the FRBs at different redshifts. The EP breaking leads to large-scale correlations between different FRBs, since the additional dispersion is sensitive to potential fluctuations rather than density, introducing an additional $k^{-2}$ factor via Poisson's equation. It is crucial to note that the mean DM-$z$ relation is unaffected.

\subsection{Forecast}
In \Cref{tab:fundamental_physics} we summarise the possible constraints on EP breaking as well as massive photons for the different stages and bands of SKA. Each single number corresponds to the $68\,\%$ confidence interval of the respective parameter. The given interval for each parameter corresponds to the most optimistic and pessimistic curve in terms of numbers of FRBs for each stage and band (compare \Cref{fig:Mid2AA4pz,fig:mag_ingal}). We assume 5 years of operations for FRB observation and marginalise over the host and Milky Way contribution, assuming a prior on the cosmological parameters from \citet{planck_collaboration_planck_2020}. The forecast for the photon mass is a couple of magnitudes better than limits from Gamma-ray bursts \citep[see e.g.][]{2021PhRvD.104j3516B} thanks to the number of FRBs. Compared to the first constraints on the photon mass \citep[see e,g,][]{2016PhLB..757..548B}, this is also a significant jump due to the larger redshifts of FRBs located and observed with SKA.

\section{Conclusions}
In this chapter we discuss the potential contributions of a population of FRBs discovered with the SKA to cosmology and fundamental physics. A companion chapter presents how the SKA-discovered FRBs can be used to study the distribution of baryons in the Universe: \citet{Caleb01.2026.SKA}. Quantitative discussions were possible thanks to a distribution of FRBs over redshift generated using the population synthesis code \texttt{DM-z}. The fact that FRBs were discussed in chapters from several different SKA Science Working Groups demonstrates the broad interest in FRBs as cosmological probes. 

We found that with a catalog of FRBs detected with the AA4 configuration, the magnetic fields in intervening galaxies (Section~\ref{sec:mag_gal}) could be measured. The SKA may detect $\sim$1 FRB per year that is lensed by a foreground galaxy or cluster. Such systems can be used, in principle, to measure cosmological parameters if the lensed images can also be detected, which will likely require dedicated follow-up by additional telescopes. The microlensing signature in FRBs can be used to constrain the density of, for example, primordial black holes or MACHOs, and due to the deliterious effects of scattering on the identification of FRB undergoing lensing, the highest bands may give the SKA an advantage for this science case. In terms of fundamental physics, the catalog of SKA-discovered FRBs may provide a limit to the photon mass that is a few orders of magnitude better than what can be done with gamma-ray bursts. 

In summary, the key strengths of the SKA are its sensitivity and frequency coverage. Sensitivity is critical for detecting sources out to large redshifts. While there are a few competing surveys coming online at similar frequencies as SKA-mid band 1 and band 2, we find that SKA-low will be peerless at its observing range and may find the majority of SKA-discovered FRBs. Given that propagation effects are highly frequency dependent, the low-frequency insight will be important for precise measurements of these effects in individual sources, as well as provide key insights into the systematics impacting statistical inference from large samples of FRBs. 

\textbf{Dedicated to the memory of J.P. Macquart, who wrote the original SKA FRB Science Book chapter “Fast Transients at Cosmological Distances with the SKA”.}

\section{Acknowledgments}

AS thanks Yik Ki (Jackie) Ma for several useful discussions. AS acknowledges support from the Australian Research Council's Discovery Early Career Researcher Award (DECRA, project~DE250100003) and the Australia-Germany Joint Research Cooperation Scheme of Universities Australia (UA--DAAD, 2025--2026). LGS is a Lise Meitner Research Group leader and acknowledges funding from the Max Planck Society. D.M. acknowledges support from the French government under the France 2030 investment plan, as part of the Initiative d'Excellence d'Aix-Marseille Universit\'e -- A*MIDEX (AMX-23-CEI-088). JJ-S acknowledges support through Australian Research Council Discovery Project DP220102305.

\bibliographystyle{abbrvnat-maxbibnames4}
\bibliography{chapter} % if your bibtex file is called example.bib

@incollection{Caleb01.2026.SKA, author = {Manisha Caleb and author2 and author3 and author4 and author5},title = {},year = {2026},publisher = {},note = {arXiv search: Report number AASKAII/Caleb01},booktitle = {Advancing Astrophysics with the SKA -- II (AASKAII)}}

@incollection{Curtin01.2026.SKA, author = {Alice P. Curtin and author2 and author3 and author4 and author5},title = {},year = {2026},publisher = {},note = {arXiv search: Report number AASKAII/Curtin01},booktitle = {Advancing Astrophysics with the SKA -- II (AASKAII)}}

@ARTICLE{BURSTT,
       author = {{Lin}, Hsiu-Hsien and {Lin}, Kai-yang and {Li}, Chao-Te and {Tseng}, Yao-Huan and {Jiang}, Homin and {Wang}, Jen-Hung and {Cheng}, Jen-Chieh and {Pen}, Ue-Li and {Chen}, Ming-Tang and {Chen}, Pisin and {Chen}, Yaocheng and {Goto}, Tomotsugu and {Hashimoto}, Tetsuya and {Hwang}, Yuh-Jing and {King}, Sun-Kun and {Kubo}, Derek and {Kuo}, Chung-Yun and {Mills}, Adam and {Nam}, Jiwoo and {Oshiro}, Peter and {Shen}, Chang-Shao and {Tseng}, Hsien-Chun and {Wang}, Shih-Hao and {Wu}, Vigo Feng-Shun and {Bower}, Geoffrey and {Chang}, Shu-Hao and {Chen}, Pai-An and {Chen}, Ying-Chih and {Chiang}, Yi-Kuan and {Fedynitch}, Anatoli and {Gusinskaia}, Nina and {Ho}, Simon C. -C. and {Hsiao}, Tiger Y. -Y. and {Hu}, Chin-Ping and {Huang}, Yau De and {J{\'a}uregui Garc{\'\i}a}, Jos{\'e} Miguel and {Kim}, Seong Jin and {Kuo}, Cheng-Yu and {Ling}, Decmend Fang-Jie and {On}, Alvina Y.~L. and {Peterson}, Jeffrey B. and {R. Raquel}, Bjorn Jasper and {Su}, Shih-Chieh and {Uno}, Yuri and {Wu}, Cossas K. -W. and {Yamasaki}, Shotaro and {Zhu}, Hong-Ming},
        title = "{BURSTT: Bustling Universe Radio Survey Telescope in Taiwan}",
      journal = {\pasp},
         year = 2022,
        month = sep,
       volume = {134},
       number = {1039},
          eid = {094106},
        pages = {094106},
          doi = {10.1088/1538-3873/ac8f71},
archivePrefix = {arXiv},
       eprint = {2206.08983},
 primaryClass = {astro-ph.IM},
       adsurl = {https://ui.adsabs.harvard.edu/abs/2022PASP..134i4106L}
}

@ARTICLE{crafts21,
       author = {{Niu}, Chen-Hui and {Li}, Di and {Luo}, Rui and {Wang}, Wei-Yang and {Yao}, Jumei and {Zhang}, Bing and {Zhu}, Wei-Wei and {Wang}, Pei and {Ye}, Haoyang and {Zhang}, Yong-Kun and {Niu}, Jia-rui and {Tang}, Ning-yu and {Duan}, Ran and {Krco}, Marko and {Dai}, Shi and {Feng}, Yi and {Miao}, Chenchen and {Pan}, Zhichen and {Qian}, Lei and {Xue}, Mengyao and {Yuan}, Mao and {Yue}, Youling and {Zhang}, Lei and {Zhang}, Xinxin},
        title = "{CRAFTS for Fast Radio Bursts: Extending the Dispersion-Fluence Relation with New FRBs Detected by FAST}",
      journal = {\apjl},
         year = 2021,
        month = mar,
       volume = {909},
       number = {1},
          eid = {L8},
        pages = {L8},
          doi = {10.3847/2041-8213/abe7f0},
archivePrefix = {arXiv},
       eprint = {2102.10546},
 primaryClass = {astro-ph.HE},
       adsurl = {https://ui.adsabs.harvard.edu/abs/2021ApJ...909L...8N}
}

@INPROCEEDINGS{DSA2000,
       author = {{Hallinan}, Gregg and {Ravi}, V. and {Weinreb}, S. and {Kocz}, J. and {Huang}, Y. and {Woody}, D.~P. and {Lamb}, J. and {D'Addario}, L. and {Catha}, M. and {Law}, C. and {Kulkarni}, S.~R. and {Phinney}, E.~S. and {Eastwood}, M.~W. and {Bouman}, K. and {McLaughlin}, M. and {Ransom}, S. and {Siemens}, X. and {Cordes}, J. and {Lynch}, R. and {Kaplan}, D. and {Brazier}, A. and {Bhatnagar}, S. and {Myers}, S. and {Walter}, F. and {Gaensler}, B.},
        title = "{The DSA-2000 {\textemdash} A Radio Survey Camera}",
    booktitle = {\baas},
         year = 2019,
       volume = {51},
        month = sep,
          eid = {255},
        pages = {255},
          doi = {10.48550/arXiv.1907.07648},
archivePrefix = {arXiv},
       eprint = {1907.07648},
 primaryClass = {astro-ph.IM},
       adsurl = {https://ui.adsabs.harvard.edu/abs/2019BAAS...51g.255H}
}

@INPROCEEDINGS{CHORD,
       author = {{Vanderlinde}, Keith and {Liu}, Adrian and {Gaensler}, Bryan and {Bond}, Dick and {Hinshaw}, Gary and {Ng}, Cherry and {Chiang}, Cynthia and {Stairs}, Ingrid and {Brown}, Jo-Anne and {Sievers}, Jonathan and {Mena}, Juan and {Smith}, Kendrick and {Bandura}, Kevin and {Masui}, Kiyoshi and {Spekkens}, Kristine and {Belostotski}, Leo and {Dobbs}, Matt and {Turok}, Neil and {Boyle}, Patrick and {Rupen}, Michael and {Landecker}, Tom and {Pen}, Ue-Li and {Kaspi}, Victoria},
        title = "{The Canadian Hydrogen Observatory and Radio-transient Detector (CHORD)}",
    booktitle = {Canadian Long Range Plan for Astronomy and Astrophysics White Papers},
         year = 2019,
       volume = {2020},
        month = oct,
          eid = {28},
        pages = {28},
          doi = {10.5281/zenodo.3765414},
archivePrefix = {arXiv},
       eprint = {1911.01777},
 primaryClass = {astro-ph.IM},
       adsurl = {https://ui.adsabs.harvard.edu/abs/2019clrp.2020...28V}
}

@ARTICLE{Sammons2023,
       author = {{Sammons}, Mawson W. and {Deller}, Adam T. and {Glowacki}, Marcin and {Gourdji}, Kelly and {James}, C.~W. and {Prochaska}, J. Xavier and {Qiu}, Hao and {Scott}, Danica R. and {Shannon}, R.~M. and {Trott}, C.~M.},
        title = "{Two-screen scattering in CRAFT FRBs}",
      journal = {MNRAS},
         year = 2023,
        month = nov,
       volume = {525},
       number = {4},
        pages = {5653-5668},
          doi = {10.1093/mnras/stad2631},
archivePrefix = {arXiv},
       eprint = {2305.11477},
 primaryClass = {astro-ph.HE},
       adsurl = {https://ui.adsabs.harvard.edu/abs/2023MNRAS.525.5653S}
}

@ARTICLE{2025OJAp....8E.127R,
       author = {{Reischke}, Robert and {Kova{\v{c}}}, Michael and {Nicola}, Andrina and {Hagstotz}, Steffen and {Schneider}, Aurel},
        title = "{An analytical model for the dispersion measure of Fast Radio Burst host galaxies}",
      journal = {The Open Journal of Astrophysics},
         year = 2025,
        month = sep,
       volume = {8},
          eid = {127},
        pages = {127},
          doi = {10.33232/001c.143819},
archivePrefix = {arXiv},
       eprint = {2411.17682},
 primaryClass = {astro-ph.CO},
       adsurl = {https://ui.adsabs.harvard.edu/abs/2025OJAp....8E.127R}
}

@ARTICLE{Ocker2022,
       author = {{Ocker}, Stella Koch and {Cordes}, James M. and {Chatterjee}, Shami and {Niu}, Chen-Hui and {Li}, Di and {McKee}, James W. and {Law}, Casey J. and {Tsai}, Chao-Wei and {Anna-Thomas}, Reshma and {Yao}, Ju-Mei and {Cruces}, Marilyn},
        title = "{The Large Dispersion and Scattering of FRB 20190520B Are Dominated by the Host Galaxy}",
      journal = {\apj},
         year = 2022,
        month = jun,
       volume = {931},
       number = {2},
          eid = {87},
        pages = {87},
          doi = {10.3847/1538-4357/ac6504},
archivePrefix = {arXiv},
       eprint = {2202.13458},
 primaryClass = {astro-ph.HE},
       adsurl = {https://ui.adsabs.harvard.edu/abs/2022ApJ...931...87O}
}

@ARTICLE{Pradeep2025,
       author = {{Pradeep E.~T.}, Sachin and {Sprenger}, Tim and {Wucknitz}, Olaf and {Main}, Robert A. and {Spitler}, Laura G.},
        title = "{Scintillometry of fast radio bursts: Resolution effects in two-screen models}",
      journal = {\aap},
         year = 2025,
        month = aug,
       volume = {700},
          eid = {A99},
        pages = {A99},
          doi = {10.1051/0004-6361/202554202},
archivePrefix = {arXiv},
       eprint = {2505.04576},
 primaryClass = {astro-ph.GA},
       adsurl = {https://ui.adsabs.harvard.edu/abs/2025A&A...700A..99P}
}

@BOOK{KleinF2015,
       author = {{Klein}, U. and {Fletcher}, A.},
        title = "{Galactic and Intergalactic Magnetic Fields}",
         year = 2015,
       adsurl = {https://ui.adsabs.harvard.edu/abs/2015gimf.book.....K}
}

@BOOK{ShukurovS2021,
       author = {{Shukurov}, A.~M. and {Subramanian}, Kandaswamy},
        title = "{Astrophysical Magnetic Fields: From Galaxies to the Early Universe}",
         year = 2021,
          doi = {10.1017/9781139046657},
       adsurl = {https://ui.adsabs.harvard.edu/abs/2021amff.book.....S}
}

@ARTICLE{KulsrudEA1997,
       author = {{Kulsrud}, Russell M. and {Cen}, Renyue and {Ostriker}, Jeremiah P. and {Ryu}, Dongsu},
        title = "{The Protogalactic Origin for Cosmic Magnetic Fields}",
      journal = {\apj},
         year = 1997,
        month = may,
       volume = {480},
       number = {2},
        pages = {481-491},
          doi = {10.1086/303987},
archivePrefix = {arXiv},
       eprint = {astro-ph/9607141},
 primaryClass = {astro-ph},
       adsurl = {https://ui.adsabs.harvard.edu/abs/1997ApJ...480..481K}
}

@ARTICLE{Subramanian2016,
       author = {{Subramanian}, Kandaswamy},
        title = "{The origin, evolution and signatures of primordial magnetic fields}",
      journal = {Reports on Progress in Physics},
         year = 2016,
        month = jul,
       volume = {79},
       number = {7},
          eid = {076901},
        pages = {076901},
          doi = {10.1088/0034-4885/79/7/076901},
archivePrefix = {arXiv},
       eprint = {1504.02311},
 primaryClass = {astro-ph.CO},
       adsurl = {https://ui.adsabs.harvard.edu/abs/2016RPPh...79g6901S}
}

@ARTICLE{SetaF2020,
       author = {{Seta}, Amit and {Federrath}, Christoph},
        title = "{Seed magnetic fields in turbulent small-scale dynamos}",
      journal = {\mnras},
         year = 2020,
        month = dec,
       volume = {499},
       number = {2},
        pages = {2076-2086},
          doi = {10.1093/mnras/staa2978},
archivePrefix = {arXiv},
       eprint = {2009.12024},
 primaryClass = {astro-ph.GA},
       adsurl = {https://ui.adsabs.harvard.edu/abs/2020MNRAS.499.2076S}
}

@ARTICLE{Beck2015,
       author = {{Beck}, Rainer},
        title = "{Magnetic fields in spiral galaxies}",
      journal = {\aapr},
         year = 2015,
        month = dec,
       volume = {24},
          eid = {4},
        pages = {4},
          doi = {10.1007/s00159-015-0084-4},
archivePrefix = {arXiv},
       eprint = {1509.04522},
 primaryClass = {astro-ph.GA},
       adsurl = {https://ui.adsabs.harvard.edu/abs/2015A&ARv..24....4B}
}

@INPROCEEDINGS{Haverkorn2015,
       author = {{Haverkorn}, Marijke},
        title = "{Magnetic Fields in the Milky Way}",
    booktitle = {Magnetic Fields in Diffuse Media},
         year = 2015,
       editor = {{Lazarian}, Alexander and {de Gouveia Dal Pino}, Elisabete M. and {Melioli}, Claudio},
       series = {\apssl},
       volume = {407},
        month = jan,
        pages = {483},
          doi = {10.1007/978-3-662-44625-6_17},
archivePrefix = {arXiv},
       eprint = {1406.0283},
 primaryClass = {astro-ph.GA},
       adsurl = {https://ui.adsabs.harvard.edu/abs/2015ASSL..407..483H}
}

@ARTICLE{SetaMCG2025,
       author = {{Seta}, Amit and {McClure-Griffiths}, N.~M.},
        title = "{Magnetic fields in the multiphase interstellar medium of the Milky Way: turbulent kinetic and magnetic energy density relation}",
      journal = {\mnras},
         year = 2025,
        month = may,
       volume = {539},
       number = {2},
        pages = {1024-1039},
          doi = {10.1093/mnras/staf520},
archivePrefix = {arXiv},
       eprint = {2503.23634},
 primaryClass = {astro-ph.GA},
       adsurl = {https://ui.adsabs.harvard.edu/abs/2025MNRAS.539.1024S}
}

@BOOK{RuzmaikinEA1988,
       author = {{Ruzmaikin}, Aleksandr A. and {Sokolov}, Dmitrii D. and {Shukurov}, Anvar M.},
        title = "{Magnetic Fields of Galaxies}",
         year = 1988,
       volume = {133},
          doi = {10.1007/978-94-009-2835-0},
       adsurl = {https://ui.adsabs.harvard.edu/abs/1988ASSL..133.....R}
}

@ARTICLE{BrandenburgS2005,
       author = {{Brandenburg}, Axel and {Subramanian}, Kandaswamy},
        title = "{Astrophysical magnetic fields and nonlinear dynamo theory}",
      journal = {\physrep},
         year = 2005,
        month = oct,
       volume = {417},
       number = {1-4},
        pages = {1-209},
          doi = {10.1016/j.physrep.2005.06.005},
archivePrefix = {arXiv},
       eprint = {astro-ph/0405052},
 primaryClass = {astro-ph},
       adsurl = {https://ui.adsabs.harvard.edu/abs/2005PhR...417....1B}
}

@ARTICLE{Rincon2019,
       author = {{Rincon}, Fran{\c{c}}ois},
        title = "{Dynamo theories}",
      journal = {Journal of Plasma Physics},
         year = 2019,
        month = aug,
       volume = {85},
       number = {4},
          eid = {205850401},
        pages = {205850401},
          doi = {10.1017/S0022377819000539},
archivePrefix = {arXiv},
       eprint = {1903.07829},
 primaryClass = {physics.plasm-ph},
       adsurl = {https://ui.adsabs.harvard.edu/abs/2019JPlPh..85d2001R}
}

@ARTICLE{SchekochihinEA2004,
       author = {{Schekochihin}, Alexander A. and {Cowley}, Steven C. and {Taylor}, Samuel F. and {Maron}, Jason L. and {McWilliams}, James C.},
        title = "{Simulations of the Small-Scale Turbulent Dynamo}",
      journal = {\apj},
         year = 2004,
        month = sep,
       volume = {612},
       number = {1},
        pages = {276-307},
          doi = {10.1086/422547},
archivePrefix = {arXiv},
       eprint = {astro-ph/0312046},
 primaryClass = {astro-ph},
       adsurl = {https://ui.adsabs.harvard.edu/abs/2004ApJ...612..276S}
}

@ARTICLE{HaugenEA2004,
       author = {{Haugen}, Nils Erland and {Brandenburg}, Axel and {Dobler}, Wolfgang},
        title = "{Simulations of nonhelical hydromagnetic turbulence}",
      journal = {\pre},
         year = 2004,
        month = jul,
       volume = {70},
       number = {1},
          eid = {016308},
        pages = {016308},
          doi = {10.1103/PhysRevE.70.016308},
archivePrefix = {arXiv},
       eprint = {astro-ph/0307059},
 primaryClass = {astro-ph},
       adsurl = {https://ui.adsabs.harvard.edu/abs/2004PhRvE..70a6308H}
}

@ARTICLE{SetaEA2020,
       author = {{Seta}, Amit and {Bushby}, Paul J. and {Shukurov}, Anvar and {Wood}, Toby S.},
        title = "{Saturation mechanism of the fluctuation dynamo at Pr$_{M}$ {\ensuremath{\geq}} 1}",
      journal = {Physical Review Fluids},
         year = 2020,
        month = apr,
       volume = {5},
       number = {4},
          eid = {043702},
        pages = {043702},
          doi = {10.1103/PhysRevFluids.5.043702},
archivePrefix = {arXiv},
       eprint = {2003.07997},
 primaryClass = {astro-ph.GA},
       adsurl = {https://ui.adsabs.harvard.edu/abs/2020PhRvF...5d3702S}
}

@ARTICLE{GentEA2013,
       author = {{Gent}, F.~A. and {Shukurov}, A. and {Sarson}, G.~R. and {Fletcher}, A. and {Mantere}, M.~J.},
        title = "{The supernova-regulated ISM - II. The mean magnetic field.}",
      journal = {\mnras},
         year = 2013,
        month = mar,
       volume = {430},
        pages = {L40-L44},
          doi = {10.1093/mnrasl/sls042},
archivePrefix = {arXiv},
       eprint = {1206.6784},
 primaryClass = {astro-ph.GA},
       adsurl = {https://ui.adsabs.harvard.edu/abs/2013MNRAS.430L..40G}
}

@ARTICLE{SetaF2021,
       author = {{Seta}, Amit and {Federrath}, Christoph},
        title = "{Saturation mechanism of the fluctuation dynamo in supersonic turbulent plasmas}",
      journal = {Physical Review Fluids},
         year = 2021,
        month = oct,
       volume = {6},
       number = {10},
          eid = {103701},
        pages = {103701},
          doi = {10.1103/PhysRevFluids.6.103701},
archivePrefix = {arXiv},
       eprint = {2109.11698},
 primaryClass = {astro-ph.GA},
       adsurl = {https://ui.adsabs.harvard.edu/abs/2021PhRvF...6j3701S}
}

@ARTICLE{GentEA2023,
       author = {{Gent}, Frederick A. and {Mac Low}, Mordecai-Mark and {Korpi-Lagg}, Maarit J. and {Singh}, Nishant K.},
        title = "{The Small-scale Dynamo in a Multiphase Supernova-driven Medium}",
      journal = {\apj},
         year = 2023,
        month = feb,
       volume = {943},
       number = {2},
          eid = {176},
        pages = {176},
          doi = {10.3847/1538-4357/acac20},
archivePrefix = {arXiv},
       eprint = {2210.04460},
 primaryClass = {astro-ph.GA},
       adsurl = {https://ui.adsabs.harvard.edu/abs/2023ApJ...943..176G}
}

@ARTICLE{FarnesEA2014,
       author = {{Farnes}, J.~S. and {O'Sullivan}, S.~P. and {Corrigan}, M.~E. and {Gaensler}, B.~M.},
        title = "{Faraday Rotation from Magnesium II Absorbers toward Polarized Background Radio Sources}",
      journal = {\apj},
         year = 2014,
        month = nov,
       volume = {795},
       number = {1},
          eid = {63},
        pages = {63},
          doi = {10.1088/0004-637X/795/1/63},
archivePrefix = {arXiv},
       eprint = {1406.2526},
 primaryClass = {astro-ph.GA},
       adsurl = {https://ui.adsabs.harvard.edu/abs/2014ApJ...795...63F}
}

@ARTICLE{ShahS2021,
       author = {{Shah}, Hilay and {Seta}, Amit},
        title = "{Magnetic fields in elliptical galaxies: using the Laing-Garrington effect in radio galaxies and polarized emission from background radio sources}",
      journal = {\mnras},
         year = 2021,
        month = nov,
       volume = {508},
       number = {1},
        pages = {1371-1388},
          doi = {10.1093/mnras/stab2500},
archivePrefix = {arXiv},
       eprint = {2108.12793},
 primaryClass = {astro-ph.GA},
       adsurl = {https://ui.adsabs.harvard.edu/abs/2021MNRAS.508.1371S}
}

@ARTICLE{AmaralEA2021,
       author = {{Amaral}, A.~D. and {Vernstrom}, T. and {Gaensler}, B.~M.},
        title = "{Constraints on large-scale magnetic fields in the intergalactic medium using cross-correlation methods}",
      journal = {\mnras},
         year = 2021,
        month = may,
       volume = {503},
       number = {2},
        pages = {2913-2926},
          doi = {10.1093/mnras/stab564},
archivePrefix = {arXiv},
       eprint = {2102.11312},
 primaryClass = {astro-ph.CO},
       adsurl = {https://ui.adsabs.harvard.edu/abs/2021MNRAS.503.2913A}
}

@ARTICLE{MalikEA2020,
       author = {{Malik}, Sunil and {Chand}, Hum and {Seshadri}, T.~R.},
        title = "{Role of Intervening Mg II Absorbers on the Rotation Measure and Fractional Polarization of the Background Quasars}",
      journal = {\apj},
         year = 2020,
        month = feb,
       volume = {890},
       number = {2},
          eid = {132},
        pages = {132},
          doi = {10.3847/1538-4357/ab6bd5},
archivePrefix = {arXiv},
       eprint = {1710.10396},
 primaryClass = {astro-ph.GA},
       adsurl = {https://ui.adsabs.harvard.edu/abs/2020ApJ...890..132M}
}

@ARTICLE{BernetEA2008,
       author = {{Bernet}, Martin L. and {Miniati}, Francesco and {Lilly}, Simon J. and {Kronberg}, Philipp P. and {Dessauges-Zavadsky}, Miroslava},
        title = "{Strong magnetic fields in normal galaxies at high redshift}",
      journal = {\nat},
         year = 2008,
        month = jul,
       volume = {454},
       number = {7202},
        pages = {302-304},
          doi = {10.1038/nature07105},
archivePrefix = {arXiv},
       eprint = {0807.3347},
 primaryClass = {astro-ph},
       adsurl = {https://ui.adsabs.harvard.edu/abs/2008Natur.454..302B}
}

@ARTICLE{MaoEA2017,
       author = {{Mao}, S.~A. and {Carilli}, C. and {Gaensler}, B.~M. and {Wucknitz}, O. and {Keeton}, C. and {Basu}, A. and {Beck}, R. and {Kronberg}, P.~P. and {Zweibel}, E.},
        title = "{Detection of microgauss coherent magnetic fields in a galaxy five billion years ago}",
      journal = {Nature Astronomy},
         year = 2017,
        month = aug,
       volume = {1},
        pages = {621-626},
          doi = {10.1038/s41550-017-0218-x},
archivePrefix = {arXiv},
       eprint = {1708.07844},
 primaryClass = {astro-ph.GA},
       adsurl = {https://ui.adsabs.harvard.edu/abs/2017NatAs...1..621M}
}

@ARTICLE{KovacsEA2025,
       author = {{Kovacs}, Timea Orsolya and {Mao}, Sui Ann and {Basu}, Aritra and {Ma}, Yik Ki and {Gaensler}, B.~M.},
        title = "{The halo magnetic field of a spiral galaxy at z=0.414}",
      journal = {arXiv e-prints},
         year = 2025,
        month = jul,
          eid = {arXiv:2507.12542},
        pages = {arXiv:2507.12542},
          doi = {10.48550/arXiv.2507.12542},
archivePrefix = {arXiv},
       eprint = {2507.12542},
 primaryClass = {astro-ph.GA},
       adsurl = {https://ui.adsabs.harvard.edu/abs/2025arXiv250712542K}
}

@ARTICLE{PakmorEA2020,
       author = {{Pakmor}, R{\"u}diger and {van de Voort}, Freeke and {Bieri}, Rebekka and {G{\'o}mez}, Facundo A. and {Grand}, Robert J.~J. and {Guillet}, Thomas and {Marinacci}, Federico and {Pfrommer}, Christoph and {Simpson}, Christine M. and {Springel}, Volker},
        title = "{Magnetizing the circumgalactic medium of disc galaxies}",
      journal = {\mnras},
         year = 2020,
        month = nov,
       volume = {498},
       number = {3},
        pages = {3125-3137},
          doi = {10.1093/mnras/staa2530},
archivePrefix = {arXiv},
       eprint = {1911.11163},
 primaryClass = {astro-ph.GA},
       adsurl = {https://ui.adsabs.harvard.edu/abs/2020MNRAS.498.3125P}
}

@ARTICLE{vanEA2021,
       author = {{van de Voort}, Freeke and {Bieri}, Rebekka and {Pakmor}, R{\"u}diger and {G{\'o}mez}, Facundo A. and {Grand}, Robert J.~J. and {Marinacci}, Federico},
        title = "{The effect of magnetic fields on properties of the circumgalactic medium}",
      journal = {\mnras},
         year = 2021,
        month = mar,
       volume = {501},
       number = {4},
        pages = {4888-4902},
          doi = {10.1093/mnras/staa3938},
archivePrefix = {arXiv},
       eprint = {2008.07537},
 primaryClass = {astro-ph.GA},
       adsurl = {https://ui.adsabs.harvard.edu/abs/2021MNRAS.501.4888V}
}

@ARTICLE{AkahoriEA2016,
       author = {{Akahori}, Takuya and {Ryu}, Dongsu and {Gaensler}, B.~M.},
        title = "{Fast Radio Bursts as Probes of Magnetic Fields in the Intergalactic Medium}",
      journal = {\apj},
         year = 2016,
        month = jun,
       volume = {824},
       number = {2},
          eid = {105},
        pages = {105},
          doi = {10.3847/0004-637X/824/2/105},
archivePrefix = {arXiv},
       eprint = {1602.03235},
 primaryClass = {astro-ph.CO},
       adsurl = {https://ui.adsabs.harvard.edu/abs/2016ApJ...824..105A}
}

@ARTICLE{HacksteinEA2019,
       author = {{Hackstein}, S. and {Br{\"u}ggen}, M. and {Vazza}, F. and {Gaensler}, B.~M. and {Heesen}, V.},
        title = "{Fast radio burst dispersion measures and rotation measures and the origin of intergalactic magnetic fields}",
      journal = {\mnras},
         year = 2019,
        month = sep,
       volume = {488},
       number = {3},
        pages = {4220-4238},
          doi = {10.1093/mnras/stz2033},
archivePrefix = {arXiv},
       eprint = {1907.09650},
 primaryClass = {astro-ph.CO},
       adsurl = {https://ui.adsabs.harvard.edu/abs/2019MNRAS.488.4220H}
}

@ARTICLE{HacksteinEA2020,
       author = {{Hackstein}, S. and {Br{\"u}ggen}, M. and {Vazza}, F. and {Rodrigues}, L.~F.~S.},
        title = "{Redshift estimates for fast radio bursts and implications on intergalactic magnetic fields}",
      journal = {\mnras},
         year = 2020,
        month = nov,
       volume = {498},
       number = {4},
        pages = {4811-4829},
          doi = {10.1093/mnras/staa2572},
archivePrefix = {arXiv},
       eprint = {2008.10536},
 primaryClass = {astro-ph.CO},
       adsurl = {https://ui.adsabs.harvard.edu/abs/2020MNRAS.498.4811H}
}

@ARTICLE{KovacsEA2024,
       author = {{Kovacs}, Timea Orsolya and {Mao}, Sui Ann and {Basu}, Aritra and {Ma}, Yik Ki and {Pakmor}, Ruediger and {Spitler}, Laura G. and {Walker}, Charles R.~H.},
        title = "{Dispersion and rotation measures from fast radio burst (FRB) host galaxies based on the TNG50 simulation}",
      journal = {\aap},
         year = 2024,
        month = oct,
       volume = {690},
          eid = {A47},
        pages = {A47},
          doi = {10.1051/0004-6361/202347459},
archivePrefix = {arXiv},
       eprint = {2407.16748},
 primaryClass = {astro-ph.GA},
       adsurl = {https://ui.adsabs.harvard.edu/abs/2024A&A...690A..47K}
}

@ARTICLE{ShahEA2025,
       author = {{Shah}, Hilay and {van de Voort}, Freeke and {Seta}, Amit and {Federrath}, Christoph},
        title = "{Understanding gas mixing in the circumgalactic medium}",
      journal = {\mnras},
         year = 2025,
        month = aug,
       volume = {541},
       number = {3},
        pages = {2471-2492},
          doi = {10.1093/mnras/staf1066},
archivePrefix = {arXiv},
       eprint = {2505.21980},
 primaryClass = {astro-ph.GA},
       adsurl = {https://ui.adsabs.harvard.edu/abs/2025MNRAS.541.2471S}
}

@ARTICLE{SetaF2024,
       author = {{Seta}, Amit and {Federrath}, Christoph},
        title = "{Structure functions with higher-order stencils as a probe to separate small- and large-scale magnetic fields}",
      journal = {\mnras},
         year = 2024,
        month = sep,
       volume = {533},
       number = {2},
        pages = {1875-1886},
          doi = {10.1093/mnras/stae1935},
archivePrefix = {arXiv},
       eprint = {2408.04156},
 primaryClass = {astro-ph.GA},
       adsurl = {https://ui.adsabs.harvard.edu/abs/2024MNRAS.533.1875S}
}

@ARTICLE{PadmanabhanEA2023,
       author = {{Padmanabhan}, Hamsa and {Loeb}, Abraham},
        title = "{A New Limit on Intergalactic Magnetic Fields on Subkiloparsec Scales from Fast Radio Bursts}",
      journal = {\apjl},
         year = 2023,
        month = mar,
       volume = {946},
       number = {1},
          eid = {L18},
        pages = {L18},
          doi = {10.3847/2041-8213/acc3a1},
archivePrefix = {arXiv},
       eprint = {2301.08259},
 primaryClass = {astro-ph.HE},
       adsurl = {https://ui.adsabs.harvard.edu/abs/2023ApJ...946L..18P}
}

@ARTICLE{RaviEA2016,
       author = {{Ravi}, V. and {Shannon}, R.~M. and {Bailes}, M. and {Bannister}, K. and {Bhandari}, S. and {Bhat}, N.~D.~R. and {Burke-Spolaor}, S. and {Caleb}, M. and {Flynn}, C. and {Jameson}, A. and {Johnston}, S. and {Keane}, E.~F. and {Kerr}, M. and {Tiburzi}, C. and {Tuntsov}, A.~V. and {Vedantham}, H.~K.},
        title = "{The magnetic field and turbulence of the cosmic web measured using a brilliant fast radio burst}",
      journal = {Science},
         year = 2016,
        month = dec,
       volume = {354},
       number = {6317},
        pages = {1249-1252},
          doi = {10.1126/science.aaf6807},
archivePrefix = {arXiv},
       eprint = {1611.05758},
 primaryClass = {astro-ph.HE},
       adsurl = {https://ui.adsabs.harvard.edu/abs/2016Sci...354.1249R}
}

@ARTICLE{DurrerN2013,
       author = {{Durrer}, Ruth and {Neronov}, Andrii},
        title = "{Cosmological magnetic fields: their generation, evolution and observation}",
      journal = {\aapr},
         year = 2013,
        month = jun,
       volume = {21},
          eid = {62},
        pages = {62},
          doi = {10.1007/s00159-013-0062-7},
archivePrefix = {arXiv},
       eprint = {1303.7121},
 primaryClass = {astro-ph.CO},
       adsurl = {https://ui.adsabs.harvard.edu/abs/2013A&ARv..21...62D}
}

@ARTICLE{WidrowEA2012,
       author = {{Widrow}, Lawrence M. and {Ryu}, Dongsu and {Schleicher}, Dominik R.~G. and {Subramanian}, Kandaswamy and {Tsagas}, Christos G. and {Treumann}, Rudolf A.},
        title = "{The First Magnetic Fields}",
      journal = {\ssr},
         year = 2012,
        month = may,
       volume = {166},
       number = {1-4},
        pages = {37-70},
          doi = {10.1007/s11214-011-9833-5},
archivePrefix = {arXiv},
       eprint = {1109.4052},
 primaryClass = {astro-ph.CO},
       adsurl = {https://ui.adsabs.harvard.edu/abs/2012SSRv..166...37W}
}

@ARTICLE{BrentjensB2005,
       author = {{Brentjens}, M.~A. and {de Bruyn}, A.~G.},
        title = "{Faraday rotation measure synthesis}",
      journal = {\aap},
         year = 2005,
        month = oct,
       volume = {441},
       number = {3},
        pages = {1217-1228},
          doi = {10.1051/0004-6361:20052990},
archivePrefix = {arXiv},
       eprint = {astro-ph/0507349},
 primaryClass = {astro-ph},
       adsurl = {https://ui.adsabs.harvard.edu/abs/2005A&A...441.1217B}
}

@ARTICLE{MillerB2013,
       author = {{Miller}, Matthew J. and {Bregman}, Joel N.},
        title = "{The Structure of the Milky Way's Hot Gas Halo}",
      journal = {\apj},
         year = 2013,
        month = jun,
       volume = {770},
       number = {2},
          eid = {118},
        pages = {118},
          doi = {10.1088/0004-637X/770/2/118},
archivePrefix = {arXiv},
       eprint = {1305.2430},
 primaryClass = {astro-ph.GA},
       adsurl = {https://ui.adsabs.harvard.edu/abs/2013ApJ...770..118M}
}

@ARTICLE{HussainiEA2025,
       author = {{Hussaini}, Maryam and {Connor}, Liam and {Konietzka}, Ralf M. and {Ravi}, Vikram and {Faber}, Jakob and {Sharma}, Kritti and {Sherman}, Myles},
        title = "{A Correlation Between FRB Dispersion Measure and Foreground Large-Scale Structure}",
      journal = {arXiv e-prints},
         year = 2025,
        month = jun,
          eid = {arXiv:2506.04186},
        pages = {arXiv:2506.04186},
          doi = {10.48550/arXiv.2506.04186},
archivePrefix = {arXiv},
       eprint = {2506.04186},
 primaryClass = {astro-ph.CO},
       adsurl = {https://ui.adsabs.harvard.edu/abs/2025arXiv250604186H}
}

@ARTICLE{Ocker2025CGMScattering,
       author = {{Ocker}, Stella Koch and {Chen}, Mandy C. and {Oh}, S. Peng and {Sharma}, Prateek},
        title = "{Microphysics of Circumgalactic Turbulence Probed by Fast Radio Bursts and Quasars}",
      journal = {\apj},
         year = 2025,
        month = jul,
       volume = {988},
       number = {1},
          eid = {69},
        pages = {69},
          doi = {10.3847/1538-4357/ade0bc},
archivePrefix = {arXiv},
       eprint = {2503.02329},
 primaryClass = {astro-ph.GA},
       adsurl = {https://ui.adsabs.harvard.edu/abs/2025ApJ...988...69O}
}

@article{Petroff2019,
 archiveprefix = {arXiv},
 author = {Petroff, E. and Hessels, J. W. T. and Lorimer, D. R.},
 doi = {10.1007/s00159-019-0116-6},
 eid = {4},
 eprint = {1904.07947},
 journal = {\aapr},
 month = {May},
 number = {1},
 pages = {4},
 primaryclass = {astro-ph.HE},
 title = {Fast radio bursts},
 url = {https://ui.adsabs.harvard.edu/abs/2019A&ARv..27....4P},
 volume = {27},
 year = {2019}
}

@Article{CHIME2022,
  author        = {{CHIME Collaboration} and Amiri, Mandana and Bandura, Kevin and Boskovic, Anja and Chen, Tianyue and Cliche, Jean-Fran{\c{c}}ois and Deng, Meiling and Denman, Nolan and Dobbs, Matt and Fandino, Mateus and Foreman, Simon and Halpern, Mark and Hanna, David and Hill, Alex S. and Hinshaw, Gary and H{\"o}fer, Carolin and Kania, Joseph and Klages, Peter and Landecker, T. L. and MacEachern, Joshua and Masui, Kiyoshi and Mena-Parra, Juan and Milutinovic, Nikola and Mirhosseini, Arash and Newburgh, Laura and Nitsche, Rick and Ordog, Anna and Pen, Ue-Li and Pinsonneault-Marotte, Tristan and Polzin, Ava and Reda, Alex and Renard, Andre and Shaw, J. Richard and Siegel, Seth R. and Singh, Saurabh and Smegal, Rick and Tretyakov, Ian and van Gassen, Kwinten and Vanderlinde, Keith and Wang, Haochen and Wiebe, Donald V. and Willis, James S. and Wulf, Dallas},
  journal       = {\apjs},
  title         = {An Overview of CHIME, the Canadian Hydrogen Intensity Mapping Experiment},
  year          = {2022},
  month         = aug,
  number        = {2},
  pages         = {29},
  volume        = {261},
  archiveprefix = {arXiv},
  doi           = {10.3847/1538-4365/ac6fd9},
  eid           = {29},
  eprint        = {2201.07869},
  primaryclass  = {astro-ph.IM},
  url           = {https://ui.adsabs.harvard.edu/abs/2022ApJS..261...29C},
}

@Article{Hotan2021,
  author        = {{Hotan}, A.~W. and {Bunton}, J.~D. and {Chippendale}, A.~P. and {Whiting}, M. and {Tuthill}, J. and {Moss}, V.~A. and {McConnell}, D. and {Amy}, S.~W. and {Huynh}, M.~T. and {Allison}, J.~R. and {Anderson}, C.~S. and {Bannister}, K.~W. and {Bastholm}, E. and {Beresford}, R. and {Bock}, D.~C. -J. and {Bolton}, R. and {Chapman}, J.~M. and {Chow}, K. and {Collier}, J.~D. and {Cooray}, F.~R. and {Cornwell}, T.~J. and {Diamond}, P.~J. and {Edwards}, P.~G. and {Feain}, I.~J. and {Franzen}, T.~M.~O. and {George}, D. and {Gupta}, N. and {Hampson}, G.~A. and {Harvey-Smith}, L. and {Hayman}, D.~B. and {Heywood}, I. and {Jacka}, C. and {Jackson}, C.~A. and {Jackson}, S. and {Jeganathan}, K. and {Johnston}, S. and {Kesteven}, M. and {Kleiner}, D. and {Koribalski}, B.~S. and {Lee-Waddell}, K. and {Lenc}, E. and {Lensson}, E.~S. and {Mackay}, S. and {Mahony}, E.~K. and {McClure-Griffiths}, N.~M. and {McConigley}, R. and {Mirtschin}, P. and {Ng}, A.~K. and {Norris}, R.~P. and {Pearce}, S.~E. and {Phillips}, C. and {Pilawa}, M.~A. and {Raja}, W. and {Reynolds}, J.~E. and {Roberts}, P. and {Roxby}, D.~N. and {Sadler}, E.~M. and {Shields}, M. and {Schinckel}, A.~E.~T. and {Serra}, P. and {Shaw}, R.~D. and {Sweetnam}, T. and {Troup}, E.~R. and {Tzioumis}, A. and {Voronkov}, M.~A. and {Westmeier}, T.},
  journal       = {\pasa},
  title         = {{Australian square kilometre array pathfinder: I. system description}},
  year          = {2021},
  month         = {March},
  pages         = {e009},
  volume        = {38},
  adsurl        = {https://ui.adsabs.harvard.edu/abs/2021PASA...38....9H},
  archiveprefix = {arXiv},
  doi           = {10.1017/pasa.2021.1},
  eid           = {e009},
  eprint        = {2102.01870},
  primaryclass  = {astro-ph.IM},
}

@Article{Shannon2024,
  author        = {Shannon, R.~M. and Bannister, K.~W. and Bera, A. and Bhandari, S. and Day, C.~K. and Deller, A.~T. and Dial, T. and Dobie, D. and Ekers, R.~D. and Fong, W. f. and Glowacki, M. and Gordon, A.~C. and Gourdji, K. and Jaini, A. and James, C.~W. and Kumar, P. and Mahony, E.~K. and Marnoch, L. and Muller, A.~R. and Prochaska, J.~X. and Qiu, H. and Ryder, S.~D. and Sadler, E.~M. and Scott, D.~R. and Tejos, N. and Uttarkar, P.~A. and Wang, Y.},
  journal       = {arXiv e-prints},
  title         = {The Commensal Real-time ASKAP Fast Transient incoherent-sum survey},
  year          = {2024},
  month         = aug,
  pages         = {arXiv:2408.02083},
  archiveprefix = {arXiv},
  doi           = {10.48550/arXiv.2408.02083},
  eid           = {arXiv:2408.02083},
  eprint        = {2408.02083},
  primaryclass  = {astro-ph.HE},
  url           = {https://ui.adsabs.harvard.edu/abs/2024arXiv240802083S},
}

@Article{CHIME2021,
  author        = {{CHIME/FRB Collaboration} and {Amiri}, Mandana and {Andersen}, Bridget C. and {Bandura}, Kevin and {Berger}, Sabrina and {Bhardwaj}, Mohit and {Boyce}, Michelle M. and {Boyle}, P.~J. and {Brar}, Charanjot and {Breitman}, Daniela and {Cassanelli}, Tomas and {Chawla}, Pragya and {Chen}, Tianyue and {Cliche}, J. -F. and {Cook}, Amanda and {Cubranic}, Davor and {Curtin}, Alice P. and {Deng}, Meiling and {Dobbs}, Matt and {Dong}, Fengqiu Adam and {Eadie}, Gwendolyn and {Fandino}, Mateus and {Fonseca}, Emmanuel and {Gaensler}, B.~M. and {Giri}, Utkarsh and {Good}, Deborah C. and {Halpern}, Mark and {Hill}, Alex S. and {Hinshaw}, Gary and {Josephy}, Alexander and {Kaczmarek}, Jane F. and {Kader}, Zarif and {Kania}, Joseph W. and {Kaspi}, Victoria M. and {Landecker}, T.~L. and {Lang}, Dustin and {Leung}, Calvin and {Li}, Dongzi and {Lin}, Hsiu-Hsien and {Masui}, Kiyoshi W. and {McKinven}, Ryan and {Mena-Parra}, Juan and {Merryfield}, Marcus and {Meyers}, Bradley W. and {Michilli}, Daniele and {Milutinovic}, Nikola and {Mirhosseini}, Arash and {M{\"u}nchmeyer}, Moritz and {Naidu}, Arun and {Newburgh}, Laura and {Ng}, Cherry and {Patel}, Chitrang and {Pen}, Ue-Li and {Petroff}, Emily and {Pinsonneault-Marotte}, Tristan and {Pleunis}, Ziggy and {Rafiei-Ravandi}, Masoud and {Rahman}, Mubdi and {Ransom}, Scott M. and {Renard}, Andre and {Sanghavi}, Pranav and {Scholz}, Paul and {Shaw}, J. Richard and {Shin}, Kaitlyn and {Siegel}, Seth R. and {Sikora}, Andrew E. and {Singh}, Saurabh and {Smith}, Kendrick M. and {Stairs}, Ingrid and {Tan}, Chia Min and {Tendulkar}, S.~P. and {Vanderlinde}, Keith and {Wang}, Haochen and {Wulf}, Dallas and {Zwaniga}, A.~V.},
  journal       = {\apjs},
  title         = {{The First CHIME/FRB Fast Radio Burst Catalog}},
  year          = {2021},
  month         = {December},
  number        = {2},
  pages         = {59},
  volume        = {257},
  adsurl        = {https://ui.adsabs.harvard.edu/abs/2021ApJS..257...59C},
  archiveprefix = {arXiv},
  doi           = {10.3847/1538-4365/ac33ab},
  eid           = {59},
  eprint        = {2106.04352},
  primaryclass  = {astro-ph.HE},
}

@Article{Law2023,
  author        = {Law, C. J. and Sharma, K. and Ravi, V. and Chen, G. and Catha, M. and Connor, L. and Faber, J. T. and Hallinan, G. and Harnach, C. and Hellbourg, G. and Hobbs, R. and Hodge, D. and Hodges, M. and Lamb, J. W. and Rasmussen, P. and Sherman, M. B. and Shi, J. and Simard, D. and Squillace, R. and Weinreb, S. and Woody, D. P. and Yadlapalli, N.},
  journal       = {arXiv e-prints},
  title         = {Deep Synoptic Array Science: First FRB and Host Galaxy Catalog},
  year          = {2023},
  month         = jul,
  pages         = {arXiv:2307.03344},
  archiveprefix = {arXiv},
  doi           = {10.48550/arXiv.2307.03344},
  eid           = {arXiv:2307.03344},
  eprint        = {2307.03344},
  primaryclass  = {astro-ph.HE},
  url           = {https://ui.adsabs.harvard.edu/abs/2023arXiv230703344L},
}

@ARTICLE{2020AcA....70...87J,
       author = {{Jaroszy{\'n}ski}, M.},
        title = "{FRBs: the Dispersion Measure of Host Galaxies}",
      journal = {\actaa},
         year = 2020,
        month = jun,
       volume = {70},
       number = {2},
        pages = {87-100},
          doi = {10.32023/0001-5237/70.2.1},
archivePrefix = {arXiv},
       eprint = {2008.04634},
 primaryClass = {astro-ph.GA},
       adsurl = {https://ui.adsabs.harvard.edu/abs/2020AcA....70...87J}
}

@article{mcquinn_locating_2014,
	title = {Locating the "{Missing}" {Baryons} with {Extragalactic} {Dispersion} {Measure} {Estimates}},
	volume = {780},
	issn = {0004-637X},
	url = {http://adsabs.harvard.edu/abs/2014ApJ...780L..33M},
	doi = {10.1088/2041-8205/780/2/L33},
	urldate = {2020-06-01},
	journal = {\apjl},
	author = {McQuinn, Matthew},
	month = jan,
	year = {2014},
	pages = {L33},
}

@ARTICLE{2025arXiv250707991K,
              author = {{Kova{\v{c}}}, Michael and {Nicola}, Andrina and {Bucko}, Jozef and {Schneider}, Aurel and {Reischke}, Robert and {Giri}, Sambit K. and {Teyssier}, Romain and {Schaller}, Matthieu and {Schaye}, Joop},
        title = "{Baryonification II: constraining feedback with X-ray and kinematic Sunyaev-Zel'dovich observations}",
      journal = {\jcap},
         year = 2025,
        month = nov,
       volume = {2025},
       number = {11},
          eid = {046},
        pages = {046},
          doi = {10.1088/1475-7516/2025/11/046},
archivePrefix = {arXiv},
       eprint = {2507.07991},
 primaryClass = {astro-ph.CO},
       adsurl = {https://ui.adsabs.harvard.edu/abs/2025JCAP...11..046K}
}

@ARTICLE{2022MNRAS.512..285R,
       author = {{Reischke}, Robert and {Hagstotz}, Steffen and {Lilow}, Robert},
        title = "{Consistent equivalence principle tests with fast radio bursts}",
      journal = {\mnras},
         year = 2022,
        month = may,
       volume = {512},
       number = {1},
        pages = {285-290},
          doi = {10.1093/mnras/stab3571},
archivePrefix = {arXiv},
       eprint = {2102.11554},
 primaryClass = {astro-ph.CO},
       adsurl = {https://ui.adsabs.harvard.edu/abs/2022MNRAS.512..285R}
}

@ARTICLE{2017PhRvD..95l3010S,
       author = {{Shao}, Lijing and {Zhang}, Bing},
        title = "{Bayesian framework to constrain the photon mass with a catalog of fast radio bursts}",
      journal = {\prd},
         year = 2017,
        month = jun,
       volume = {95},
       number = {12},
          eid = {123010},
        pages = {123010},
          doi = {10.1103/PhysRevD.95.123010},
archivePrefix = {arXiv},
       eprint = {1705.01278},
 primaryClass = {hep-ph},
       adsurl = {https://ui.adsabs.harvard.edu/abs/2017PhRvD..95l3010S}
}

@ARTICLE{2016PhLB..757..548B,
       author = {{Bonetti}, Luca and {Ellis}, John and {Mavromatos}, Nikolaos E. and {Sakharov}, Alexander S. and {Sarkisyan-Grinbaum}, Edward K. and {Spallicci}, Alessandro D.~A.~M.},
        title = "{Photon mass limits from fast radio bursts}",
      journal = {Physics Letters B},
         year = 2016,
        month = jun,
       volume = {757},
        pages = {548-552},
          doi = {10.1016/j.physletb.2016.04.035},
       adsurl = {https://ui.adsabs.harvard.edu/abs/2016PhLB..757..548B}
}

@ARTICLE{2021PhRvD.104j3516B,
       author = {{Bartlett}, D.~J. and {Desmond}, H. and {Ferreira}, P.~G. and {Jasche}, J.},
        title = "{Constraints on quantum gravity and the photon mass from gamma ray bursts}",
      journal = {\prd},
         year = 2021,
        month = nov,
       volume = {104},
       number = {10},
          eid = {103516},
        pages = {103516},
          doi = {10.1103/PhysRevD.104.103516},
archivePrefix = {arXiv},
       eprint = {2109.07850},
 primaryClass = {gr-qc},
       adsurl = {https://ui.adsabs.harvard.edu/abs/2021PhRvD.104j3516B}
}

@ARTICLE{2023MNRAS.523.6264R,
       author = {{Reischke}, Robert and {Hagstotz}, Steffen},
        title = "{Consistent constraints on the equivalence principle from localized fast radio bursts}",
      journal = {\mnras},
         year = 2023,
        month = aug,
       volume = {523},
       number = {4},
        pages = {6264-6271},
          doi = {10.1093/mnras/stad1866},
archivePrefix = {arXiv},
       eprint = {2302.10072},
 primaryClass = {astro-ph.CO},
       adsurl = {https://ui.adsabs.harvard.edu/abs/2023MNRAS.523.6264R}
}

@ARTICLE{2025PASA...42...17H,
       author = {{Hoffmann}, Jordan Luke and {James}, Clancy and {Glowacki}, Marcin and {Prochaska}, Xavier and {Gordon}, Alexa and {Deller}, Adam and {Shannon}, Ryan M. and {Ryder}, Stuart},
        title = "{Modelling DSA, FAST, and CRAFT surveys in a z-DM analysis and constraining a minimum FRB energy}",
      journal = {\pasa},
         year = 2025,
        month = jan,
       volume = {42},
          eid = {e017},
        pages = {e017},
          doi = {10.1017/pasa.2024.127},
archivePrefix = {arXiv},
       eprint = {2408.04878},
 primaryClass = {astro-ph.CO},
       adsurl = {https://ui.adsabs.harvard.edu/abs/2025PASA...42...17H}
}

@ARTICLE{2014ARA&A..52..415M,
       author = {{Madau}, Piero and {Dickinson}, Mark},
        title = "{Cosmic Star-Formation History}",
      journal = {\araa},
         year = 2014,
        month = aug,
       volume = {52},
        pages = {415-486},
          doi = {10.1146/annurev-astro-081811-125615},
archivePrefix = {arXiv},
       eprint = {1403.0007},
 primaryClass = {astro-ph.CO},
       adsurl = {https://ui.adsabs.harvard.edu/abs/2014ARA&A..52..415M}
}

@ARTICLE{2016PhR...643....1M,
       author = {{Marsh}, David J.~E.},
        title = "{Axion cosmology}",
      journal = {\physrep},
         year = 2016,
        month = jul,
       volume = {643},
        pages = {1-79},
          doi = {10.1016/j.physrep.2016.06.005},
archivePrefix = {arXiv},
       eprint = {1510.07633},
 primaryClass = {astro-ph.CO},
       adsurl = {https://ui.adsabs.harvard.edu/abs/2016PhR...643....1M}
}

@ARTICLE{2022MNRAS.511..662H,
       author = {{Hagstotz}, Steffen and {Reischke}, Robert and {Lilow}, Robert},
        title = "{A new measurement of the Hubble constant using fast radio bursts}",
      journal = {\mnras},
         year = 2022,
        month = mar,
       volume = {511},
       number = {1},
        pages = {662-667},
          doi = {10.1093/mnras/stac077},
archivePrefix = {arXiv},
       eprint = {2104.04538},
 primaryClass = {astro-ph.CO},
       adsurl = {https://ui.adsabs.harvard.edu/abs/2022MNRAS.511..662H}
}

@ARTICLE{2023ApJ...946...58C,
       author = {{Cook}, Amanda M. and {Bhardwaj}, Mohit and {Gaensler}, B.~M. and {Scholz}, Paul and {Eadie}, Gwendolyn M. and {Hill}, Alex S. and {Kaspi}, Victoria M. and {Masui}, Kiyoshi W. and {Curtin}, Alice P. and {Dong}, Fengqiu Adam and {Fonseca}, Emmanuel and {Herrera-Martin}, Antonio and {Kaczmarek}, Jane and {Lanman}, Adam E. and {Lazda}, Mattias and {Leung}, Calvin and {Meyers}, Bradley W. and {Michilli}, Daniele and {Pandhi}, Ayush and {Pearlman}, Aaron B. and {Pleunis}, Ziggy and {Ransom}, Scott and {Rahman}, Mubdi and {Sand}, Ketan R. and {Shin}, Kaitlyn and {Smith}, Kendrick and {Stairs}, Ingrid and {Stenning}, David C.},
        title = "{An FRB Sent Me a DM: Constraining the Electron Column of the Milky Way Halo with Fast Radio Burst Dispersion Measures from CHIME/FRB}",
      journal = {\apj},
         year = 2023,
        month = apr,
       volume = {946},
       number = {2},
          eid = {58},
        pages = {58},
          doi = {10.3847/1538-4357/acbbd0},
archivePrefix = {arXiv},
       eprint = {2301.03502},
 primaryClass = {astro-ph.GA},
       adsurl = {https://ui.adsabs.harvard.edu/abs/2023ApJ...946...58C}
}

@ARTICLE{2020ApJ...888..105Y,
       author = {{Yamasaki}, Shotaro and {Totani}, Tomonori},
        title = "{The Galactic Halo Contribution to the Dispersion Measure of Extragalactic Fast Radio Bursts}",
      journal = {\apj},
         year = 2020,
        month = jan,
       volume = {888},
       number = {2},
          eid = {105},
        pages = {105},
          doi = {10.3847/1538-4357/ab58c4},
archivePrefix = {arXiv},
       eprint = {1909.00849},
 primaryClass = {astro-ph.HE},
       adsurl = {https://ui.adsabs.harvard.edu/abs/2020ApJ...888..105Y}
}

@article{caleb_2023_subarcsec,
    author = {Caleb, M and Driessen, L N and Gordon, A C and Tejos, N and Bernales, L and Qiu, H and Chibueze, J O and Stappers, B W and Rajwade, K M and Cavallaro, F and Wang, Y and Kumar, P and Majid, W A and Wharton, R S and Naudet, C J and Bezuidenhout, M C and Jankowski, F and Malenta, M and Morello, V and Sanidas, S and Surnis, M P and Barr, E D and Chen, W and Kramer, M and Fong, W and Kilpatrick, C D and Prochaska, J Xavier and Simha, S and Venter, C and Heywood, I and Kundu, A and Schussler, F},
    title = {A subarcsec localized fast radio burst with a significant host galaxy dispersion measure contribution},
    journal = {\mnras},
    volume = {524},
    number = {2},
    pages = {2064-2077},
    year = {2023},
    month = {06},
    issn = {0035-8711},
    doi = {10.1093/mnras/stad1839},
    url = {https://doi.org/10.1093/mnras/stad1839},
    eprint = {https://academic.oup.com/mnras/article-pdf/524/2/2064/53672010/stad1839.pdf},
}

@ARTICLE{Rajwade+22,
       author = {{Rajwade}, K.~M. and {Bezuidenhout}, M.~C. and {Caleb}, M. and {Driessen}, L.~N. and {Jankowski}, F. and {Malenta}, M. and {Morello}, V. and {Sanidas}, S. and {Stappers}, B.~W. and {Surnis}, M.~P. and {Barr}, E.~D. and {Chen}, W. and {Kramer}, M. and {Wu}, J. and {Buchner}, S. and {Serylak}, M. and {Combes}, F. and {Fong}, W. and {Gupta}, N. and {Jagannathan}, P. and {Kilpatrick}, C.~D. and {Krogager}, J.-K. and {Noterdaeme}, P. and {N{\'u}nẽz}, C. and {Prochaska}, J. Xavier and {Srianand}, R. and {Tejos}, N.},
        title = "{First discoveries and localizations of Fast Radio Bursts with MeerTRAP: real-time, commensal MeerKAT survey}",
      journal = {\mnras},
         year = 2022,
        month = aug,
       volume = {514},
       number = {2},
        pages = {1961-1974},
          doi = {10.1093/mnras/stac1450},
archivePrefix = {arXiv},
       eprint = {2205.14600},
 primaryClass = {astro-ph.HE},
       adsurl = {https://ui.adsabs.harvard.edu/abs/2022MNRAS.514.1961R}
}

@ARTICLE{2025arXiv250801648C,
       author = {{Caleb}, Manisha and {Nanayakkara}, Themiya and {Stappers}, Benjamin and {Pastor-Marazuela}, In{\'e}s and {Khrykin}, Ilya S. and {Glazebrook}, Karl and {Tejos}, Nicolas and {Prochaska}, J. Xavier and {Rajwade}, Kaustubh and {Mas-Ribas}, Lluis and {Driessen}, Laura N. and {Fong}, Wen-fai and {Gordon}, Alexa C. and {Hoffmann}, Jordan and {James}, Clancy W. and {Jankowski}, Fabian and {Kahinga}, Lordrick and {Kramer}, Michael and {Simha}, Sunil and {Barr}, Ewan D. and {Christiaan Bezuidenhout}, Mechiel and {Deng}, Xihan and {Lin}, Zeren and {Marnoch}, Lachlan and {Martin}, Christopher D. and {Nugent}, Anya and {Shaji}, Kavya and {Tian}, Jun},
        title = "{A fast radio burst from the first 3 billion years of the Universe}",
      journal = {arXiv e-prints},
         year = 2025,
        month = aug,
          eid = {arXiv:2508.01648},
        pages = {arXiv:2508.01648},
          doi = {10.48550/arXiv.2508.01648},
archivePrefix = {arXiv},
       eprint = {2508.01648},
 primaryClass = {astro-ph.HE},
       adsurl = {https://ui.adsabs.harvard.edu/abs/2025arXiv250801648C}
}

@article{chime_2021_first,
doi = {10.3847/1538-4365/ac33ab},
url = {https://dx.doi.org/10.3847/1538-4365/ac33ab},
year = {2021},
month = {dec},
publisher = {The American Astronomical Society},
volume = {257},
number = {2},
pages = {59},
author = {{CHIME/FRB Collaboration} and Mandana Amiri and Bridget C. Andersen and Kevin Bandura and Sabrina Berger and Mohit Bhardwaj and Michelle M. Boyce and P. J. Boyle and Charanjot Brar and Daniela Breitman and Tomas Cassanelli and Pragya Chawla and Tianyue Chen and J.-F. Cliche and Amanda Cook and Davor Cubranic and Alice P. Curtin and Meiling Deng and Matt Dobbs and Fengqiu (Adam) Dong and Gwendolyn Eadie and Mateus Fandino and Emmanuel Fonseca and B. M. Gaensler and Utkarsh Giri and Deborah C. Good and Mark Halpern and Alex S. Hill and Gary Hinshaw and Alexander Josephy and Jane F. Kaczmarek and Zarif Kader and Joseph W. Kania and Victoria M. Kaspi and T. L. Landecker and Dustin Lang and Calvin Leung and Dongzi Li and Hsiu-Hsien Lin and Kiyoshi W. Masui and Ryan Mckinven and Juan Mena-Parra and Marcus Merryfield and Bradley W. Meyers and Daniele Michilli and Nikola Milutinovic and Arash Mirhosseini and Moritz Münchmeyer and Arun Naidu and Laura Newburgh and Cherry Ng and Chitrang Patel and Ue-Li Pen and Emily Petroff and Tristan Pinsonneault-Marotte and Ziggy Pleunis and Masoud Rafiei-Ravandi and Mubdi Rahman and Scott M. Ransom and Andre Renard and Pranav Sanghavi and Paul Scholz and J. Richard Shaw and Kaitlyn Shin and Seth R. Siegel and Andrew E. Sikora and Saurabh Singh and Kendrick M. Smith and Ingrid Stairs and Chia Min Tan and S. P. Tendulkar and Keith Vanderlinde and Haochen Wang and Dallas Wulf and A. V. Zwaniga},
title = {The First CHIME/FRB Fast Radio Burst Catalog},
journal = {\apjs}
}

@ARTICLE{2017ApJ...835...29Y,
       author = {{Yao}, J.~M. and {Manchester}, R.~N. and {Wang}, N.},
        title = "{A New Electron-density Model for Estimation of Pulsar and FRB Distances}",
      journal = {\apj},
         year = 2017,
        month = jan,
       volume = {835},
       number = {1},
          eid = {29},
        pages = {29},
          doi = {10.3847/1538-4357/835/1/29},
archivePrefix = {arXiv},
       eprint = {1610.09448},
 primaryClass = {astro-ph.GA},
       adsurl = {https://ui.adsabs.harvard.edu/abs/2017ApJ...835...29Y}
}

@ARTICLE{2020MNRAS.496L.106K,
       author = {{Keating}, Laura C. and {Pen}, Ue-Li},
        title = "{Exploring the dispersion measure of the Milky Way halo}",
      journal = {\mnras},
         year = 2020,
        month = jul,
       volume = {496},
       number = {1},
        pages = {L106-L110},
          doi = {10.1093/mnrasl/slaa095},
archivePrefix = {arXiv},
       eprint = {2001.11105},
 primaryClass = {astro-ph.GA},
       adsurl = {https://ui.adsabs.harvard.edu/abs/2020MNRAS.496L.106K}
}

@ARTICLE{2019MNRAS.485..648P,
       author = {{Prochaska}, J. Xavier and {Zheng}, Yong},
        title = "{Probing Galactic haloes with fast radio bursts}",
      journal = {\mnras},
         year = 2019,
        month = may,
       volume = {485},
       number = {1},
        pages = {648-665},
          doi = {10.1093/mnras/stz261},
archivePrefix = {arXiv},
       eprint = {1901.11051},
 primaryClass = {astro-ph.GA},
       adsurl = {https://ui.adsabs.harvard.edu/abs/2019MNRAS.485..648P}
}

@ARTICLE{2024RNAAS...8...17O,
       author = {{Ocker}, Stella Koch and {Cordes}, James M.},
        title = "{NE2001p: A Native Python Implementation of the NE2001 Galactic Electron Density Model}",
      journal = {Research Notes of the American Astronomical Society},
         year = 2024,
        month = jan,
       volume = {8},
       number = {1},
          eid = {17},
        pages = {17},
          doi = {10.3847/2515-5172/ad1bf1},
archivePrefix = {arXiv},
       eprint = {2401.05475},
 primaryClass = {astro-ph.GA},
       adsurl = {https://ui.adsabs.harvard.edu/abs/2024RNAAS...8...17O}
}

@ARTICLE{2020ApJ...897..124O,
       author = {{Ocker}, Stella Koch and {Cordes}, James M. and {Chatterjee}, Shami},
        title = "{Electron Density Structure of the Local Galactic Disk}",
      journal = {\apj},
         year = 2020,
        month = jul,
       volume = {897},
       number = {2},
          eid = {124},
        pages = {124},
          doi = {10.3847/1538-4357/ab98f9},
archivePrefix = {arXiv},
       eprint = {2004.11921},
 primaryClass = {astro-ph.GA},
       adsurl = {https://ui.adsabs.harvard.edu/abs/2020ApJ...897..124O}
}

@ARTICLE{2024arXiv240308611T,
       author = {{Theis}, Alexander and {Hagstotz}, Steffen and {Reischke}, Robert and {Weller}, Jochen},
        title = "{Galaxy dispersion measured by Fast Radio Bursts as a probe of baryonic feedback models}",
      journal = {arXiv e-prints},
         year = 2024,
        month = mar,
          eid = {arXiv:2403.08611},
        pages = {arXiv:2403.08611},
          doi = {10.48550/arXiv.2403.08611},
archivePrefix = {arXiv},
       eprint = {2403.08611},
 primaryClass = {astro-ph.CO},
       adsurl = {https://ui.adsabs.harvard.edu/abs/2024arXiv240308611T}
}

@ARTICLE{2024ApJ...967...32M,
       author = {{Medlock}, Isabel and {Nagai}, Daisuke and {Singh}, Priyanka and {Oppenheimer}, Benjamin and {Angl{\'e}s-Alc{\'a}zar}, Daniel and {Villaescusa-Navarro}, Francisco},
        title = "{Probing the Circumgalactic Medium with Fast Radio Bursts: Insights from CAMELS}",
      journal = {\apj},
         year = 2024,
        month = may,
       volume = {967},
       number = {1},
          eid = {32},
        pages = {32},
          doi = {10.3847/1538-4357/ad3070},
archivePrefix = {arXiv},
       eprint = {2403.02313},
 primaryClass = {astro-ph.GA},
       adsurl = {https://ui.adsabs.harvard.edu/abs/2024ApJ...967...32M}
}

@article{planck_collaboration_planck_2020,
	title = {Planck 2018 results. {VI}. {Cosmological} parameters},
	volume = {641},
	issn = {0004-6361},
	url = {http://adsabs.harvard.edu/abs/2020A%26A...641A...6P},
	doi = {10.1051/0004-6361/201833910},
	urldate = {2021-01-09},
	journal = {\aap},
	author = {{Planck Collaboration} and Aghanim, N. and Akrami, Y. and Ashdown, M. and Aumont, J. and Baccigalupi, C. and Ballardini, M. and Banday, A. J. and Barreiro, R. B. and Bartolo, N. and Basak, S. and Battye, R. and Benabed, K. and Bernard, J.-P. and Bersanelli, M. and Bielewicz, P. and Bock, J. J. and Bond, J. R. and Borrill, J. and Bouchet, F. R. and Boulanger, F. and Bucher, M. and Burigana, C. and Butler, R. C. and Calabrese, E. and Cardoso, J.-F. and Carron, J. and Challinor, A. and Chiang, H. C. and Chluba, J. and Colombo, L. P. L. and Combet, C. and Contreras, D. and Crill, B. P. and Cuttaia, F. and de Bernardis, P. and de Zotti, G. and Delabrouille, J. and Delouis, J.-M. and Di Valentino, E. and Diego, J. M. and Doré, O. and Douspis, M. and Ducout, A. and Dupac, X. and Dusini, S. and Efstathiou, G. and Elsner, F. and Enßlin, T. A. and Eriksen, H. K. and Fantaye, Y. and Farhang, M. and Fergusson, J. and Fernandez-Cobos, R. and Finelli, F. and Forastieri, F. and Frailis, M. and Fraisse, A. A. and Franceschi, E. and Frolov, A. and Galeotta, S. and Galli, S. and Ganga, K. and Génova-Santos, R. T. and Gerbino, M. and Ghosh, T. and González-Nuevo, J. and Górski, K. M. and Gratton, S. and Gruppuso, A. and Gudmundsson, J. E. and Hamann, J. and Handley, W. and Hansen, F. K. and Herranz, D. and Hildebrandt, S. R. and Hivon, E. and Huang, Z. and Jaffe, A. H. and Jones, W. C. and Karakci, A. and Keihänen, E. and Keskitalo, R. and Kiiveri, K. and Kim, J. and Kisner, T. S. and Knox, L. and Krachmalnicoff, N. and Kunz, M. and Kurki-Suonio, H. and Lagache, G. and Lamarre, J.-M. and Lasenby, A. and Lattanzi, M. and Lawrence, C. R. and Le Jeune, M. and Lemos, P. and Lesgourgues, J. and Levrier, F. and Lewis, A. and Liguori, M. and Lilje, P. B. and Lilley, M. and Lindholm, V. and López-Caniego, M. and Lubin, P. M. and Ma, Y.-Z. and Macías-Pérez, J. F. and Maggio, G. and Maino, D. and Mandolesi, N. and Mangilli, A. and Marcos-Caballero, A. and Maris, M. and Martin, P. G. and Martinelli, M. and Martínez-González, E. and Matarrese, S. and Mauri, N. and McEwen, J. D. and Meinhold, P. R. and Melchiorri, A. and Mennella, A. and Migliaccio, M. and Millea, M. and Mitra, S. and Miville-Deschênes, M.-A. and Molinari, D. and Montier, L. and Morgante, G. and Moss, A. and Natoli, P. and Nørgaard-Nielsen, H. U. and Pagano, L. and Paoletti, D. and Partridge, B. and Patanchon, G. and Peiris, H. V. and Perrotta, F. and Pettorino, V. and Piacentini, F. and Polastri, L. and Polenta, G. and Puget, J.-L. and Rachen, J. P. and Reinecke, M. and Remazeilles, M. and Renzi, A. and Rocha, G. and Rosset, C. and Roudier, G. and Rubiño-Martín, J. A. and Ruiz-Granados, B. and Salvati, L. and Sandri, M. and Savelainen, M. and Scott, D. and Shellard, E. P. S. and Sirignano, C. and Sirri, G. and Spencer, L. D. and Sunyaev, R. and Suur-Uski, A.-S. and Tauber, J. A. and Tavagnacco, D. and Tenti, M. and Toffolatti, L. and Tomasi, M. and Trombetti, T. and Valenziano, L. and Valiviita, J. and Van Tent, B. and Vibert, L. and Vielva, P. and Villa, F. and Vittorio, N. and Wandelt, B. D. and Wehus, I. K. and White, M. and White, S. D. M. and Zacchei, A. and Zonca, A.},
	month = sep,
	year = {2020},
	pages = {A6},
}

@ARTICLE{macquart_census_2020,
       author = {{Macquart}, J. -P. and {Prochaska}, J.~X. and {McQuinn}, M. and {Bannister}, K.~W. and {Bhandari}, S. and {Day}, C.~K. and {Deller}, A.~T. and {Ekers}, R.~D. and {James}, C.~W. and {Marnoch}, L. and {Os{l}owski}, S. and {Phillips}, C. and {Ryder}, S.~D. and {Scott}, D.~R. and {Shannon}, R.~M. and {Tejos}, N.},
        title = "{A census of baryons in the Universe from localized fast radio bursts}",
      journal = {nat},
         year = 2020,
        month = may,
       volume = {581},
       number = {7809},
        pages = {391-395},
          doi = {10.1038/s41586-020-2300-2},
archivePrefix = {arXiv},
       eprint = {2005.13161},
 primaryClass = {astro-ph.CO},
       adsurl = {https://ui.adsabs.harvard.edu/abs/2020Natur.581..391M}
}

@ARTICLE{walters_future_2018,
       author = {{Walters}, Anthony and {Weltman}, Amanda and {Gaensler}, B.~M. and {Ma}, Yin-Zhe and {Witzemann}, Amadeus},
        title = "{Future Cosmological Constraints From Fast Radio Bursts}",
      journal = {\apj},
         year = 2018,
        month = mar,
       volume = {856},
       number = {1},
          eid = {65},
        pages = {65},
          doi = {10.3847/1538-4357/aaaf6b},
archivePrefix = {arXiv},
       eprint = {1711.11277},
 primaryClass = {astro-ph.CO},
       adsurl = {https://ui.adsabs.harvard.edu/abs/2018ApJ...856...65W}
}

@ARTICLE{walters_probing_2019,
       author = {{Walters}, Anthony and {Ma}, Yin-Zhe and {Sievers}, Jonathan and {Weltman}, Amanda},
        title = "{Probing diffuse gas with fast radio bursts}",
      journal = {\prd},
         year = 2019,
        month = nov,
       volume = {100},
       number = {10},
          eid = {103519},
        pages = {103519},
          doi = {10.1103/PhysRevD.100.103519},
archivePrefix = {arXiv},
       eprint = {1909.02821},
 primaryClass = {astro-ph.CO},
       adsurl = {https://ui.adsabs.harvard.edu/abs/2019PhRvD.100j3519W}
}

@ARTICLE{2019PhRvD.100j4047M,
       author = {{Minazzoli}, Olivier and {Johnson-McDaniel}, Nathan K. and {Sakellariadou}, Mairi},
        title = "{Shortcomings of Shapiro delay-based tests of the equivalence principle on cosmological scales}",
      journal = {\prd},
         year = 2019,
        month = nov,
       volume = {100},
       number = {10},
          eid = {104047},
        pages = {104047},
          doi = {10.1103/PhysRevD.100.104047},
archivePrefix = {arXiv},
       eprint = {1907.12453},
 primaryClass = {gr-qc},
       adsurl = {https://ui.adsabs.harvard.edu/abs/2019PhRvD.100j4047M}
}

@ARTICLE{2008JCAP...01..031J,
       author = {{Jacob}, Uri and {Piran}, Tsvi},
        title = "{Lorentz-violation-induced arrival delays of cosmological particles}",
      journal = {\jcap},
         year = 2008,
        month = jan,
       volume = {2008},
       number = {1},
          eid = {031},
        pages = {031},
          doi = {10.1088/1475-7516/2008/01/031},
archivePrefix = {arXiv},
       eprint = {0712.2170},
 primaryClass = {astro-ph},
       adsurl = {https://ui.adsabs.harvard.edu/abs/2008JCAP...01..031J}
}

@ARTICLE{Bianco2022,
       author = {{Bianco}, Federica B. and {Ivezi{\'c}}, {\v{Z}}eljko and {Jones}, R. Lynne and {Graham}, Melissa L. and {Marshall}, Phil and {Saha}, Abhijit and {Strauss}, Michael A. and {Yoachim}, Peter and {Ribeiro}, Tiago and {Anguita}, Timo and {Bauer}, A.~E. and {Bauer}, Franz E. and {Bellm}, Eric C. and {Blum}, Robert D. and {Brandt}, William N. and {Brough}, Sarah and {Catelan}, M{\'a}rcio and {Clarkson}, William I. and {Connolly}, Andrew J. and {Gawiser}, Eric and {Gizis}, John E. and {Hlo{\v{z}}ek}, Ren{\'e}e and {Kaviraj}, Sugata and {Liu}, Charles T. and {Lochner}, Michelle and {Mahabal}, Ashish A. and {Mandelbaum}, Rachel and {McGehee}, Peregrine and {Neilsen}, Jr., Eric H. and {Olsen}, Knut A.~G. and {Peiris}, Hiranya V. and {Rhodes}, Jason and {Richards}, Gordon T. and {Ridgway}, Stephen and {Schwamb}, Megan E. and {Scolnic}, Dan and {Shemmer}, Ohad and {Slater}, Colin T. and {Slosar}, An{\v{z}}e and {Smartt}, Stephen J. and {Strader}, Jay and {Street}, Rachel and {Trilling}, David E. and {Verma}, Aprajita and {Vivas}, A.~K. and {Wechsler}, Risa H. and {Willman}, Beth},
        title = "{Optimization of the Observing Cadence for the Rubin Observatory Legacy Survey of Space and Time: A Pioneering Process of Community-focused Experimental Design}",
      journal = {\apjs},
         year = 2022,
        month = jan,
       volume = {258},
       number = {1},
          eid = {1},
        pages = {1},
          doi = {10.3847/1538-4365/ac3e72},
archivePrefix = {arXiv},
       eprint = {2108.01683},
 primaryClass = {astro-ph.IM},
       adsurl = {https://ui.adsabs.harvard.edu/abs/2022ApJS..258....1B}
}

@article{Cole2000,
 adsurl = {https://ui.adsabs.harvard.edu/abs/2000MNRAS.319..168C},
 archiveprefix = {arXiv},
 author = {{Cole}, Shaun and {Lacey}, Cedric G. and {Baugh}, Carlton M. and {Frenk}, Carlos S.},
 doi = {10.1046/j.1365-8711.2000.03879.x},
 eprint = {astro-ph/0007281},
 journal = {\mnras},
 month = {November},
 number = {1},
 pages = {168-204},
 primaryclass = {astro-ph},
 title = {{Hierarchical galaxy formation}},
 volume = {319},
 year = {2000}
}

@article{Baugh2019,
 adsurl = {https://ui.adsabs.harvard.edu/abs/2019MNRAS.483.4922B},
 archiveprefix = {arXiv},
 author = {{Baugh}, C.~M. and {Gonzalez-Perez}, Violeta and {Lagos}, Claudia del P and {Lacey}, Cedric G. and {Helly}, John C. and {Jenkins}, Adrian and {Frenk}, Carlos S. and {Benson}, Andrew J. and {Bower}, Richard G. and {Cole}, Shaun},
 doi = {10.1093/mnras/sty3427},
 eprint = {1808.08276},
 journal = {\mnras},
 month = {March},
 number = {4},
 pages = {4922-4937},
 primaryclass = {astro-ph.GA},
 title = {{Galaxy formation in the Planck Millennium: the atomic hydrogen content of dark matter halos}},
 volume = {483},
 year = {2019}
}

@Article{Jahns2023,
  author        = {Jahns-Schindler, Joscha N. and Spitler, Laura G. and Walker, Charles R. H. and Baugh, Carlton M.},
  journal       = {\mnras},
  title         = {How limiting is optical follow-up for fast radio burst applications? Forecasts for radio and optical surveys},
  year          = {2023},
  month         = aug,
  number        = {4},
  pages         = {5006-5023},
  volume        = {523},
  archiveprefix = {arXiv},
  doi           = {10.1093/mnras/stad1659},
  eprint        = {2306.00084},
  primaryclass  = {astro-ph.HE},
  url           = {https://ui.adsabs.harvard.edu/abs/2023MNRAS.523.5006J},
}

\end{document}